\pdfoutput=1
\documentclass[preprint,5p,times,twocolumn]{elsarticle}

\makeatletter
\def\ps@pprintTitle{%
  \let\@oddhead\@empty
  \let\@evenhead\@empty
  \def\@oddfoot{\footnotesize\itshape Preprint \hfill \today}%
  \let\@evenfoot\@oddfoot
}
\makeatother
\biboptions{numbers,sort&compress}

\usepackage{booktabs}
\usepackage{graphicx}
\usepackage{caption}
\usepackage{subcaption}
\usepackage{multirow}
\usepackage{amsmath,amssymb,amsfonts}
\renewcommand{\le}{\leqslant}

\renewcommand{\ge}{\geqslant}

\usepackage[export]{adjustbox}
\usepackage{microtype}
\usepackage{xcolor}
\usepackage{listings}
\usepackage{tikz}
\usetikzlibrary{arrows.meta,patterns,calc,decorations.markings,
    shapes.geometric,positioning}
\usepackage[colorlinks=true,linkcolor=blue!70!black,citecolor=blue!70!black,
    urlcolor=blue!70!black]{hyperref}

\graphicspath{{./}}

\definecolor{codegreen}{rgb}{0,0.52,0.18}
\definecolor{codegray}{rgb}{0.45,0.45,0.45}
\definecolor{codepurple}{rgb}{0.55,0,0.75}
\definecolor{backcolour}{rgb}{0.975,0.975,0.98}

\lstdefinestyle{matlabstyle}{
    language=Matlab,
    backgroundcolor=\color{backcolour},
    commentstyle=\color{codegreen}\itshape,
    keywordstyle=\color{blue!80!black}\bfseries,
    numberstyle=\tiny\color{codegray},
    stringstyle=\color{codepurple},
    basicstyle=\ttfamily\fontsize{6.8pt}{8.3pt}\selectfont,
    breakatwhitespace=false,
    breaklines=true,
    captionpos=b,
    keepspaces=true,
    numbers=left,
    numbersep=5pt,
    showspaces=false,
    showstringspaces=false,
    showtabs=false,
    tabsize=2,
    frame=single,
    rulecolor=\color{black!20}
}
\DeclareMathOperator*{\minimize}{minimize}
\DeclareMathOperator*{\subjectto}{subject~to}

\newcommand{\diff}{\mathrm{d}}
\newcommand{\vect}[1]{\boldsymbol{#1}}
\newcommand{\mat}[1]{\mathbf{#1}}
\newcommand{\R}{\mathbb{R}}
\newcommand{\Ne}{N_{\mathrm{e}}}
\newcommand{\Nn}{N_{\mathrm{n}}}

\begin{document}

\begin{frontmatter}

\title{\texorpdfstring{\texttt{topAC99}}{topAC99}: A compact MATLAB implementation of
filter-free Allen--Cahn topology optimization with matrix-free field transfer}

\author[1,2,3]{Run Du\corref{cor1}}
\ead{rdu@swjtu.edu.cn}
\cortext[cor1]{Corresponding author.}

\affiliation[1]{organization={School of Mechanical Engineering,
    Southwest Jiaotong University},
    city={Chengdu},
    postcode={610031},
    state={Sichuan},
    country={China}}

\affiliation[2]{organization={Technology and Equipment of Rail Transit Operation
    and Maintenance Key Laboratory of Sichuan Province},
    city={Chengdu},
    postcode={610031},
    state={Sichuan},
    country={China}}

\affiliation[3]{organization={State Key Laboratory of Bridge Intelligent and
    Green Construction, Southwest Jiaotong University},
    city={Chengdu},
    postcode={611756},
    state={Sichuan},
    country={China}}

\begin{abstract}
This paper presents \texttt{topAC99}, a self-contained, toolbox-free 99-line
MATLAB kernel for filter-free Allen--Cahn topology optimization via the
Augmented Lagrangian method on structured quadrilateral meshes. The software architecture combines three
compact numerical components: (i)~a lumped nodal discretization with a 7-line
Kronecker-sum 5-point Laplacian that enforces homogeneous Neumann conditions
at width-invariant sparse memory cost; (ii)~matrix-free node-to-element
averaging and adjoint backpropagation via two-dimensional convolutions
(\texttt{conv2}); and (iii)~an inlined 18-line 1D-dual bisection Method of Moving
Asymptotes (MMA) solver for signed sensitivities. Benchmarks from $60\times 40$ to $300\times 200$ meshes ($121{,}002$
degrees of freedom) against convolution- and PDE-filtered SIMP baselines show
that Allen--Cahn residual and adjoint evaluations add only $0.7\%$--$1.2\%$
per-step overhead while converging in 78--92 iterations on fine grids.
Minimal-line extensions to a symmetric force inverter compliant mechanism and a
passive-deck highway arch bridge further demonstrate the versatility of the
99-line kernel.
\end{abstract}


\end{frontmatter}

\section{Introduction}\label{sec:intro}

Compact, open-source software implementations have played a foundational role in
translating computational mechanics algorithms into reproducible engineering
practice. Following Sigmund's pioneering 99-line MATLAB script~\cite{Sigmund2001}
for Solid Isotropic Material with Penalization (SIMP)~\cite{Bendsoe1988,Bendsoe2003},
a rich ecosystem of self-contained codes has accelerated research and software
prototyping across vectorized finite element assembly~\cite{Andreassen2011,Ferrari2020b},
unstructured polygonal discretizations~\cite{Talischi2012PolyTop}, three-dimensional
and freeform frameworks~\cite{LiuKai2014a,ZuoZhihao2015,Ibhadode2024a}, and
boundary-evolving level-set formulations~\cite{Challis2010a,WeiPeng2018}.

Despite this breadth, a comprehensive software survey of 122 open-source
implementations by Wang et al.~\cite{WangChao2021a} highlights a pronounced
architectural imbalance: explicit filter-based SIMP, evolutionary~\cite{HuangXiaodong2021a},
and Hamilton--Jacobi level-set codes dominate the public software landscape,
whereas compact implementations for variational diffuse-interface (phase-field)
topology optimization~\cite{Bourdin2003a,Burger2006a,Takezawa2010a,Wallin2012a}
remain remarkably scarce. Even the 115-line ``alternating active-phase'' script
by Tavakoli and Mohseni~\cite{Tavakoli2014}---sometimes conflated with
phase-field codes due to terminology and a separate Allen--Cahn study by
Tavakoli~\cite{Tavakoli2014b}---actually addresses filter-regularized
multi-material SIMP optimization via projected steepest descent without
Modica--Mortola energy or phase-field governing equations. More recently, Li et
al.~\cite{LiHao2023a} developed distributed reaction--diffusion frameworks for
high-resolution multiphysics problems, and Li et al.~\cite{LiHe2025a} introduced
a 198-line MATLAB framework (\texttt{IGA\_RDTOP}) coupling reaction--diffusion
level sets with NURBS-based isogeometric analysis. Nevertheless, on standard
bilinear quadrilateral (Q4) finite element grids, practitioners still lack a
self-contained, dependency-free reference kernel for canonical Allen--Cahn
phase-field optimization comparable in brevity and out-of-the-box portability to
\texttt{top88}~\cite{Andreassen2011} or \texttt{top99neo}~\cite{Ferrari2020b}.

Four algorithmic and software-engineering bottlenecks have historically prevented
canonical phase-field formulations from being distilled into a sub-100-line
engineering kernel:
\begin{enumerate}
    \item \textit{Operator order and transient marching}: Mass-conserving
        fourth-order Cahn--Hilliard models require mixed finite elements and
        saddle-point solvers~\cite{Wallin2012a}, whereas second-order
        Allen--Cahn gradient flows~\cite{Allen1979,Tavakoli2014b} do not conserve
        volume and rely on pseudo-time stepping with restrictive stability steps
        or nested nonlinear solves~\cite{Takezawa2010a,Sigmund2013a,Blank2017a,YuQian2021a}.
    \item \textit{Surrogate energy coupling}: Weighted objective-plus-perimeter
        surrogates~\cite{Bourdin2006a,ZhouShiwei2007a,Blank2014a} add
        Modica--Mortola interface energy directly to the structural objective, so
        changing load magnitude or objective type shifts the continuum Pareto
        stationary layout itself.
    \item \textit{Quadrature and sparse mapping overhead}: Full Galerkin
        discretizations of cubic double-well reactions require multi-point
        element Gauss quadrature loops, explicit sparse node-to-element transfer
        matrices, or auxiliary Helmholtz PDE factorizations~\cite{Lazarov2011a},
        inflating both code volume and operator setup complexity.
    \item \textit{Signed sensitivity bottleneck}: Standard SIMP compliance
        gradients are strictly non-positive and admit closed-form Optimality
        Criteria (OC) updates~\cite{Sigmund2001}. By contrast, phase-field
        restoring forces and compliant-mechanism mutual-energy
        objectives~\cite{Sigmund1997a,ZhuBenliang2020} exhibit signed nodal
        gradients, ordinarily requiring external Method of Moving Asymptotes
        (MMA) packages~\cite{Svanberg1987}.
\end{enumerate}

To decouple the diffuse-interface equilibrium manifold from mechanical objective
penalization, our companion theoretical study~\cite{DuRun2026topac} established a
continuum Allen--Cahn Residual-Constrained Topology Optimization (ACTO) framework
solved via a mass-weighted Augmented Lagrangian Method (AC-ALM), in which the
stationary conserved Allen--Cahn equation is enforced as a separate equality
constraint. However, the research-grade finite element realization
in~\cite{DuRun2026topac} relies on consistent $3\times 3$ ($3\times 3\times 3$ in
3D) Gauss quadrature passes for nonlinear reaction assembly, 9-point element-loop
Laplacian matrices, explicit distance-weighted projection operators~\cite{Guest2004a},
tight residual tolerances, and external Svanberg \texttt{mmasub} routines.

Addressing this software realization gap, the present paper introduces
\texttt{topAC99}, a standalone 99-line MATLAB implementation that reformulates
Allen--Cahn residual-constrained optimization into a compact tensor-product
architecture with matrix-free dual-grid field transfer and zero external toolbox,
MEX, or solver dependencies. Specifically, the distinct numerical and software contributions of
this work are:
\begin{enumerate}
    \item \textbf{Lumped tensor-product Allen--Cahn discretization}: Replacing
        multi-point Gauss quadrature and element diffusion loops with trapezoidal
        nodal quadrature and a 7-line Kronecker-sum 5-point discrete Laplacian
        that satisfies exact discrete zero-flux conservation ($\vect{1}^\top
        \mat{K}_{\mathrm{lap}} = \vect{0}^\top$) with a width-invariant
        $\mathcal{O}(\Nn)$ sparse memory footprint.
    \item \textbf{Matrix-free primal/adjoint dual-grid mapping}: Exploiting
        native 2D C-level convolutions (\texttt{conv2} in \texttt{'valid'} and
        \texttt{'full'} modes) to execute exact bilinear node-to-element
        averaging and transpose sensitivity backpropagation without assembling
        sparse transfer matrices, paired with a staggered continuation-weighted
        ALM update.
    \item \textbf{Inlined 18-line 1D-dual bisection MMA kernel}: Providing a
        closed-form separable convex subproblem solver for signed sensitivities
        with a conservative reciprocal volume surrogate and an analytical
        lower-bound feasibility fallback, eliminating external optimizer
        dependencies.
    \item \textbf{Quantitative software profiling and structural design
        extensibility}: Benchmarking modular runtime and memory scaling against
        variance-matched convolution- and PDE-filtered SIMP baselines, and
        demonstrating minimal-line extensions to signed mutual-energy compliant
        mechanism synthesis (a symmetric force inverter) and passive-domain
        civil bridge design.
\end{enumerate}

The remainder of this paper is organized as follows. Section~\ref{sec:theory}
concisely summarizes the governing continuum problem and derives the discrete
tensor-product operators, matrix-free dual-grid field transfer, and 1D-dual MMA solver
that distinguish \texttt{topAC99} from full-FEM implementations.
Section~\ref{sec:code} documents the modular software architecture of
\texttt{topAC99.m}. Section~\ref{sec:benchmarks} presents multi-resolution
scaling, modular runtime/memory profiling, and parameter sensitivity analyses.
Section~\ref{sec:extensions} demonstrates code extensions to compliant force
inverter synthesis and highway arch bridge design, and Section~\ref{sec:concl}
concludes the paper. The core 99-line MATLAB kernel \texttt{topAC99.m} is listed
in \ref{sec:appendix}.

\section{Discrete Numerical Realization and Matrix-Free Field Transfer}\label{sec:theory}

\subsection{Summary of the Continuum Governing Problem}\label{subsec:continuous_formulation}

Consider a bounded planar design domain $\Omega = (0, L_x)\times(0, L_y) \subset
\R^2$ discretized into $\Ne = n_x \times n_y$ unit-sized bilinear quadrilateral
(Q4) elements ($\delta x = 1$) and $\Nn = (n_x+1)(n_y+1)$ nodes. Under linear
elasticity with Dirichlet support on $\Gamma_D$ and traction on $\Gamma_t$, the
displacement field $\vect{u}$ satisfies static equilibrium with stress modulated
by a continuous nodal design field $\phi \in H^1(\Omega)$ ($\phi = 1$ solid,
$\phi = 0$ void) via standard SIMP penalization~\cite{Bendsoe2003}:
\begin{equation}\label{eq:cont_simp}
\mathbb{C}(\phi) = \bigl( E_{\min} + (E_0 - E_{\min})\phi^p \bigr)\mathbb{C}_0,
\end{equation}
with $E_0 = 1$, $E_{\min} = 10^{-9}$, Poisson's ratio $\nu = 0.3$, and $p = 3$.

To regularize topology without spatial density filters, the continuum ACTO
formulation~\cite{DuRun2026topac} imposes the Euler--Lagrange stationarity
condition of the Modica--Mortola diffuse-interface energy~\cite{Modica1977a},
equipped with the symmetric double-well potential $W(\phi) = \beta\phi^2(1-\phi)^2$
($\beta > 0$), as a pointwise equality constraint alongside the volume limit
$V_0 \in (0,1)$:
\begin{equation}\label{eq:cont_problem}
\begin{aligned}
\minimize_{\phi \in H^1(\Omega)} \quad & c(\phi) \\
\subjectto \quad & h(\phi) \equiv -\epsilon^2 \Delta \phi + W'(\phi) - \mu = 0
    \quad \text{in } \Omega, \\
& \nabla\phi \cdot \vect{n} = 0 \quad \text{on } \partial\Omega, \\
& g(\phi) = \frac{1}{|\Omega|}\int_\Omega \phi\,\diff\Omega - V_0 \le 0,
    \quad 0 \le \phi \le 1,
\end{aligned}
\end{equation}
where $c(\phi)$ is the structural objective (compliance in baseline stiffness
design or port displacement $u_{\mathrm{out}}$ in compliant mechanism synthesis)
and $W'(\phi) = \beta(2\phi - 6\phi^2 + 4\phi^3)$. Integrating $h(\phi) = 0$
over $\Omega$ under homogeneous Neumann boundary conditions determines the
nonlocal chemical potential in closed form~\cite{DuRun2026topac}:
\begin{equation}\label{eq:cont_mu}
\mu(\phi) = \frac{1}{|\Omega|} \int_\Omega W'(\phi)\,\diff\Omega,
\end{equation}
which ensures global zero mean residual over $\Omega$. Across a planar
transition layer, the stationarity condition produces a hyperbolic-tangent
equilibrium profile with correlation length $\ell_{\mathrm{AC}} = \epsilon/\sqrt{\beta}$.
Measuring the $0.1\to 0.9$ transition thickness $w$ in element units yields the
closed-form parameter mapping:
\begin{equation}\label{eq:param_conversion}
w = \sqrt{2}\ln(9)\,\ell_{\mathrm{AC}} \approx 3.11\,\ell_{\mathrm{AC}}
\implies \epsilon = \frac{w\sqrt{\beta}}{\sqrt{2}\ln(9)}.
\end{equation}
Because $\beta$ cancels out of $h(\phi) = 0$ under Eq.~\eqref{eq:param_conversion},
$w$ alone dictates geometric interface thickness, whereas $\beta$ calibrates the
magnitude of interface constraint gradients relative to $\nabla c(\phi)$.

\subsection{Lumped Tensor-Product Discretization and Width-Invariant Memory}%
\label{subsec:fe_discretization}

A central software design departure of \texttt{topAC99} from the full Galerkin
research implementation~\cite{DuRun2026topac} lies in how the weak form of
$h(\phi) = 0$,
\begin{equation}\label{eq:weak_form}
\int_\Omega \epsilon^2 \nabla v \cdot \nabla\phi\,\diff\Omega
+ \int_\Omega v\bigl(W'(\phi) - \mu\bigr)\diff\Omega = 0,
\quad \forall v \in H^1(\Omega),
\end{equation}
is discretized on structured Q4 grids. In~\cite{DuRun2026topac}, evaluating the
nonlinear reaction and tangent integrals requires a $3\times 3$ tensor-product
Gauss quadrature loop over all elements at every iteration, alongside standard
9-point FEM Laplacian assembly.

To eliminate quadrature loops and decouple the nonlinear reaction pointwise,
\texttt{topAC99} exploits the separability of bilinear Q4 shape functions
together with 1D composite trapezoidal nodal quadrature. Along each axis,
trapezoidal quadrature yields tributary weights $[0.5, 1, \dots, 1, 0.5]^\top$
in vectors $\vect{m}_x$ (length $n_x+1$) and $\vect{m}_y$ (length $n_y+1$),
whose outer product forms the 2D nodal tributary area array $\mat{M}_{\mathrm{2D}}
= \vect{m}_y \vect{m}_x^\top$ ($1$ interior, $0.5$ edge, $0.25$ corner, summing
to $|\Omega| = \Ne$). Defining the diagonal mass matrix $\mat{M} =
\mathrm{diag}(\vect{m})$ from $\vect{m} = \mathrm{vec}(\mat{M}_{\mathrm{2D}})$,
the 2D diffusion bilinear form factorizes via Fubini's theorem into a 5-point
Kronecker sum:
\begin{equation}\label{eq:kronecker_lap}
\mat{K}_{\mathrm{lap}} = \mat{D}_x \otimes \mathrm{diag}(\vect{m}_y)
+ \mathrm{diag}(\vect{m}_x) \otimes \mat{D}_y,
\end{equation}
where $\mat{D}_x$ and $\mat{D}_y$ are symmetric tridiagonal 1D Neumann
diffusion matrices (main diagonal $[1, 2, \dots, 2, 1]$ and off-diagonals
$-1$). Because every row and column of these 1D operators sums to zero,
$\mat{K}_{\mathrm{lap}}$ is symmetric positive semi-definite and satisfies
$\vect{1}^\top \mat{K}_{\mathrm{lap}} = \vect{0}^\top$ to machine precision
without boundary-node post-processing.

Under this lumped tensor-product rule, the discrete nodal residual vector
reduces to direct array arithmetic:
\begin{equation}\label{eq:discrete_h}
\vect{h}(\vect{\phi}) = \epsilon^2 \mat{K}_{\mathrm{lap}}\vect{\phi}
+ \mat{M}\bigl(W'(\vect{\phi}) - \mu(\vect{\phi})\vect{1}\bigr) = \vect{0},
\quad
\mu(\vect{\phi}) = \frac{\vect{m}^\top W'(\vect{\phi})}{|\Omega|},
\end{equation}
preserving exact discrete volume compatibility ($\vect{1}^\top\vect{h} = 0$)
for any $\vect{\phi} \in [0,1]^{\Nn}$.

From a numerical and software-engineering standpoint, Eq.~\eqref{eq:kronecker_lap}
entails two key advantages alongside a well-defined accuracy trade-off:
\begin{enumerate}
    \item \textit{Width-invariant $\mathcal{O}(\Nn)$ sparsity}: Explicit
        conical density filters~\cite{Andreassen2011} assemble a sparse matrix
        $\mat{H}_{\mathrm{flt}}$ whose stencil $(2\lceil r_{\min}\rceil - 1)^2$
        and nonzero count $\mathcal{O}(\Ne r_{\min}^2)$ grow quadratically with
        the filter radius. Conversely, $\mat{K}_{\mathrm{lap}}$ retains a fixed
        5-point stencil with at most $5\Nn$ nonzeros regardless of how wide the
        interface thickness $w$ is chosen (memory footprints are quantified in
        Section~\ref{sec:timing_breakdown}).
    \item \textit{Zero auxiliary PDE factorizations}: Implicit Helmholtz PDE
        filters~\cite{Lazarov2011a} also achieve width-independent sparsity, but
        must solve two auxiliary $\Nn\times\Nn$ linear systems per iteration (one
        for density filtering and one for adjoint sensitivity filtering). In
        \texttt{topAC99}, Eq.~\eqref{eq:discrete_h} and its adjoint require only
        forward sparse matrix-vector multiplications (SpMVs) with
        $\mat{K}_{\mathrm{lap}}$.
    \item \textit{Lumped consistency vs.\ 9-point rotational isotropy}: On a
        Cartesian grid ($\delta x = 1$), the consistent 9-point Laplacian has a
        rotationally invariant leading truncation error $-\frac{1}{12}(\delta x)^2
        \Delta^2\phi$. The 5-point lumped stencil instead yields $-\frac{1}{12}
        (\delta x)^2(\partial_x^4 + \partial_y^4)\phi$, introducing a weak 4-fold
        lattice anisotropy of order $\mathcal{O}((\delta x/\ell_{\mathrm{AC}})^2)$.
        We adopt full trapezoidal nodal quadrature to maintain variational
        consistency with the pointwise reaction $W'(\vect{\phi})$ and diagonal
        inverse $\mat{M}^{-1}$ in Eqs.~\eqref{eq:alm_continuation}--\eqref{eq:fast_Hzeta}.
        Choosing $w \ge 2.0$ elements ($\ell_{\mathrm{AC}} \ge 0.64\,\delta x$)
        resolves the $\tanh$ transition layer across multiple elements and keeps
        directional bias negligible.
\end{enumerate}

\subsection{Matrix-Free Convolution Mapping and Staggered ALM Adjoint}%
\label{subsec:alm_sensitivity}

To couple the nodal field $\vect{\phi} \in [0,1]^{\Nn}$ to element-wise
constant SIMP stiffnesses, physical element densities $\bar{\vect{\phi}} \in
[0,1]^{\Ne}$ are defined by the four-node centroid average $\bar{\vect{\phi}} =
\mat{A}\vect{\phi}$, where $\mat{A}$ is a local sub-element discretization map
whose support contracts with the mesh size ($\delta x \to 0$). Because nodes and elements form dual 2D
Cartesian arrays, \texttt{topAC99} evaluates both primal averaging and exact
adjoint sensitivity backpropagation matrix-free via native 2D convolution with
the $2\times 2$ kernel $\mat{K}_{2\times 2} = \frac{1}{4}\mat{1}_{2\times 2}$
(eliminating explicit assembly and storage of the $4\Ne$-nonzero sparse transfer matrix $\mat{A}$):
\begin{equation}\label{eq:conv2_duality}
\begin{aligned}
\bar{\vect{\phi}} &\leftrightarrow \mathtt{conv2}(\mathtt{phi},\,\mat{K}_{2\times 2},\,\mathtt{'valid'}), \\
\mat{A}^\top \frac{\partial c}{\partial \bar{\vect{\phi}}} &\leftrightarrow
\mathtt{conv2}(\mathtt{dc},\,\mat{K}_{2\times 2},\,\mathtt{'full'}).
\end{aligned}
\end{equation}
Because $\mat{A}^\top \vect{1} = \vect{m}$, the mean element volume fraction
equals the nodal quadrature volume $\vect{v}^\top\vect{\phi}$ with weight vector
$\vect{v} = \vect{m}/|\Omega|$, ensuring exact volume consistency between nodal
optimization and finite element analysis.

To enforce $\vect{h}(\vect{\phi}) = \vect{0}$ without premature interface
pinning during early topological nucleation, \texttt{topAC99} minimizes a
continuation-weighted Augmented Lagrangian functional~\cite{DuRun2026topac}:
\begin{equation}\label{eq:alm_continuation}
\mathcal{L}_{w_h}(\vect{\phi}; \vect{\lambda}_h, \rho_h) = c(\bar{\vect{\phi}})
+ w_h \left[ \vect{\lambda}_h^\top \vect{h}(\vect{\phi})
+ \frac{\rho_h}{2} \vect{h}(\vect{\phi})^\top \mat{M}^{-1}\vect{h}(\vect{\phi}) \right],
\end{equation}
where $\vect{\lambda}_h \in \R^{\Nn}$ is the intensive nodal multiplier field,
$\rho_h > 0$ is the penalty scalar, and $w_h$ ramps linearly from $0.11$ to $1$
over the first 16 iterations. The quadratic penalty induces the mesh-independent
dual RMS residual norm:
\begin{equation}\label{eq:h_norm_def}
\|\vect{h}\|_* \equiv \sqrt{\frac{1}{|\Omega|}\vect{h}^\top \mat{M}^{-1}\vect{h}}
= \sqrt{\frac{1}{|\Omega|}\sum_{i=1}^{\Nn} \frac{h_i^2}{m_i}}.
\end{equation}
Differentiating $\mathcal{L}_{w_h}$ gives $\nabla_{\vect{\phi}}\mathcal{L}_{w_h}
= \mat{A}^\top(\partial c/\partial \bar{\vect{\phi}}) + w_h \mat{H}^\top \vect{z}$
with effective multiplier $\vect{z} = \vect{\lambda}_h + \rho_h \mat{M}^{-1}\vect{h}$.
Under lumped quadrature, the non-symmetric Jacobian $\mat{H} =
\partial\vect{h}/\partial\vect{\phi}^\top$ decomposes into a symmetric operator
plus a rank-one mass-projection term~\cite{DuRun2026topac}, so its transpose
action requires only one SpMV with $\mat{K}_{\mathrm{lap}}$ and element-wise
evaluations of $W''(\phi) = \beta(2 - 12\phi + 12\phi^2)$:
\begin{equation}\label{eq:fast_Hzeta}
\mat{H}^\top \vect{z} = \epsilon^2 \mat{K}_{\mathrm{lap}}\vect{z}
+ \mat{M}\mathrm{diag}\bigl(W''(\vect{\phi})\bigr)(\vect{z} - \bar{z}\vect{1}),
\qquad
\bar{z} = \frac{\vect{m}^\top \vect{z}}{|\Omega|}.
\end{equation}
Following each accepted step, \texttt{topAC99} performs a staggered multiplier
update reusing the evaluated residual ($\vect{\lambda}_h \leftarrow
\vect{\lambda}_h + \rho_h \mat{M}^{-1}\vect{h}$), clamps entries to $[-10,10]$,
and subtracts the constant nullspace mean ($\vect{\lambda}_h \leftarrow
\vect{\lambda}_h - (\vect{v}^\top\vect{\lambda}_h)\vect{1}$, bounding
$|\lambda_{h,i}| \le 20$). Because both $\vect{\lambda}_h$ and
$\mat{M}^{-1}\vect{h}$ have zero volume-weighted mean, $\bar{z} = 0$ vanishes to
machine precision while remaining in Eq.~\eqref{eq:fast_Hzeta} to guard against
floating-point drift. Reusing $\vect{h}$ avoids an extra SpMV per step, and
convergence ($\text{ch}\to 0$) closes any lag. Whereas~\cite{DuRun2026topac}
enforces a strict asymptotic tolerance $\|\vect{h}\|_*/\beta \le 0.075$,
\texttt{topAC99} relaxes the stationarity gate to $\|\vect{h}\|_* \le 0.28\beta$
to accelerate outer convergence. Table~\ref{tab:arch_diff} contrasts this
kernel against representative SIMP codes and the full Galerkin
framework~\cite{DuRun2026topac}.

\begin{table*}[t]
\centering
\footnotesize
\caption{Software and numerical comparison across representative topology
optimization implementations.}\label{tab:arch_diff}
\setlength{\tabcolsep}{1.8pt}
\begin{tabular*}{\textwidth}{@{\extracolsep{\fill}}llll@{\extracolsep{\fill}}}
\toprule
Architectural Component & Filtered SIMP (\texttt{top88}, \texttt{top99neo}, PDE)~\cite{Andreassen2011,Ferrari2020b,Lazarov2011a} & Full-FEM AC-ALM~\cite{DuRun2026topac} & Present \texttt{topAC99} Kernel \\
\midrule
Design field \& regularity & Element-wise $\vect{x} \in L^\infty(\Omega)$ & Nodal $\vect{\phi} \in H^1(\Omega)$ (2D Q4 / 3D H8) & Nodal $\vect{\phi} \in H^1(\Omega)$ (2D tensor grid) \\
Regularization operator & Sparse $\mat{H}_{\mathrm{flt}}$, \texttt{imfilter}, or 2 PDE solves/iter & 9-pt consistent FEM $\mat{K}_{\mathrm{lap}}$ & 7-line 5-pt Kronecker $\mat{K}_{\mathrm{lap}}$ (SpMV) \\
Nonlinear reaction \& Hessian & None (or Heaviside $\tanh$ projection~\cite{Ferrari2020b}) & $3\times 3$ Gauss quadrature loops & Lumped trapezoidal pointwise arrays \\
Node--element map ($\mat{A},\mat{A}^\top$) & Element-centered ($\mat{H}_{\mathrm{flt}}$ storage $\propto r_{\min}^2$) & Sparse index gather/scatter tables & Matrix-free $2\times 2$ \texttt{conv2} kernel \\
Subproblem optimizer & OC fixed-point ($\partial c/\partial x_e \le 0$ only) & External Svanberg \texttt{mmasub} & Inlined 18-line 1D-dual bisection MMA \\
Convergence / annealing gate & Fixed move or $\beta_{\mathrm{H}}$ continuation & $\|\vect{h}\|_*/\beta \le 0.075$ & Ramped $w_h$, $\|\vect{h}\|_* \le 0.28\beta$, annealed $\delta$ \\
Code footprint \& dependencies & $88$--$110$ lines (\texttt{top99neo}: IPT/\texttt{fsparse}) & Multi-routine suite + external MMA & Exact 99 lines, base MATLAB only \\
\bottomrule
\end{tabular*}
\end{table*}

\subsection{Inlined 1D-Dual Bisection MMA Solver for Signed Sensitivities}%
\label{subsec:mma_theory}

Because both the Allen--Cahn restoring gradient $w_h\mat{H}^\top\vect{z}$ and
mutual-energy sensitivities in compliant mechanisms take positive and negative
values across $\Omega$, multiplicative OC updates are inapplicable without
ad hoc gradient clipping. Once $\vect{h}(\vect{\phi}) = \vect{0}$ is
incorporated into $\mathcal{L}_{w_h}$, Eq.~\eqref{eq:cont_problem} retains only
a single linear volume constraint, allowing \texttt{topAC99} to embed a
specialized single-constraint MMA solver~\cite{Svanberg1987} in 18 lines.

At iteration $k$, given adaptive moving asymptotes $L_i < \phi_i^k < U_i$ and
total sensitivity $df_{0,i} = (\nabla_{\vect{\phi}}\mathcal{L}_{w_h})_i$, the
objective and linear volume constraint $g(\vect{\phi}) = \vect{v}^\top\vect{\phi}
- V_0 \le 0$ are approximated by strictly convex separable surrogates:
\begin{equation}\label{eq:mma_surrogates}
\begin{aligned}
\tilde{f}_0(\vect{\phi}) &= r_0 + \sum_{i=1}^{\Nn}\left(\frac{p_{0,i}}{U_i - \phi_i}
+ \frac{q_{0,i}}{\phi_i - L_i}\right), \\
\tilde{g}(\vect{\phi}) &= r_1 + \sum_{i=1}^{\Nn}\frac{p_{1,i}}{U_i - \phi_i} \le 0,
\end{aligned}
\end{equation}
with non-negative reciprocal weights
\[
\begin{aligned}
p_{0,i} &= \bigl(\max(0,df_{0,i}) + 10^{-3}|df_{0,i}| + 10^{-6}\bigr)(U_i - \phi_i^k)^2, \\
q_{0,i} &= \bigl(\max(0,-df_{0,i}) + 10^{-3}|df_{0,i}| + 10^{-6}\bigr)(\phi_i^k - L_i)^2, \\
p_{1,i} &= v_i(U_i - \phi_i^k)^2, \qquad
r_1 = \vect{v}^\top\vect{\phi}^k - V_0 - \sum_{i=1}^{\Nn} \frac{p_{1,i}}{U_i - \phi_i^k}.
\end{aligned}
\]
By strict convexity of $(U_i - \phi_i)^{-1}$, $\tilde{g}(\vect{\phi}) \ge
g(\vect{\phi})$ for all $\vect{\phi} < \vect{U}$, so every surrogate-feasible
iterate automatically satisfies the physical volume limit.

For any dual multiplier $\lambda \ge 0$ on $\tilde{g}$, stationarity of
$\tilde{f}_0(\vect{\phi}) + \lambda\tilde{g}(\vect{\phi})$ over the safeguarded
trust-region box $[a_i, b_i]$ yields the closed-form primal map:
\begin{equation}\label{eq:mma_primal_sol}
\phi_i^*(\lambda) = \max\Biggl(a_i,\,\min\Biggl(b_i,\,
\frac{L_i\sqrt{p_{0,i} + \lambda p_{1,i}} + U_i\sqrt{q_{0,i}}}
{\sqrt{p_{0,i} + \lambda p_{1,i}} + \sqrt{q_{0,i}}}\Biggr)\Biggr),
\end{equation}
where $a_i = \max(0, \phi_i^k - \delta, L_i + 0.1(\phi_i^k - L_i))$ and
$b_i = \min(1, \phi_i^k + \delta, U_i - 0.1(U_i - \phi_i^k))$. If the
unconstrained trial $\vect{\phi}^*(0)$ violates the volume surrogate,
the monotonically decreasing scalar equation $\tilde{g}(\vect{\phi}^*(\lambda))
= 0$ is solved by 25 bisection steps on $\lambda \in [0, 10^6]$ (narrowing
the bracket by $2^{-25} \approx 3\times 10^{-8}$). Initializing the feasible
buffer to the lower box bound $\vect{a}$ provides a guaranteed analytical
fallback for active domains whenever extreme penalty gradients push the dual
root beyond $10^6$.

\section{Software Architecture and Implementation Details}\label{sec:code}

The complete 99-line function \texttt{topAC99.m} (\ref{sec:appendix}) is invoked
with six scalar arguments:
\begin{lstlisting}[numbers=none]
[c, x, loop, h_norm] = topAC99(nelx, nely, volfrac, penal, w, beta);
\end{lstlisting}
The code runs out-of-the-box in base MATLAB (R2018b+, used here for all
numerical verifications) and GNU Octave (7.0+) using only built-in array and
sparse primitives (\texttt{conv2}, \texttt{sparse}, \texttt{spdiags},
\texttt{kron}, \texttt{mldivide}). It requires neither the Image Processing
Toolbox (\texttt{imfilter}) nor external C-MEX packages
(\texttt{fsparse})~\cite{Ferrari2020b}.

\subsection{Dual-Grid Data Layout and Variable Mapping}\label{subsec:variable_mapping}

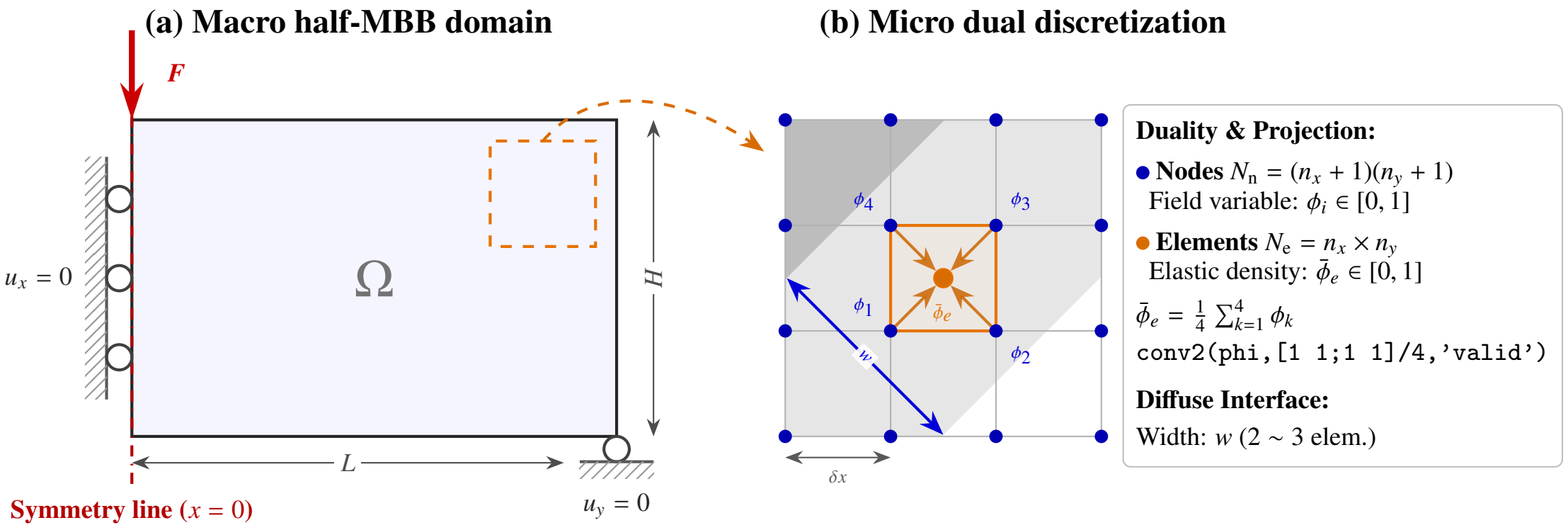
\begin{figure*}[t]
\centering
\resizebox{0.80\textwidth}{!}{%
\begin{tikzpicture}[>=Stealth, every node/.style={font=\small}]

\begin{scope}[shift={(0,0)}]
    \node[anchor=south west, font=\small\bfseries] at (0, 3.65)
        {(a) Macro half-MBB domain};

    \fill[blue!4!white] (0, 0) rectangle (4.6, 3.0);
    \draw[thick, black!85] (0, 0) rectangle (4.6, 3.0);

    \node[font=\Large\bfseries, text=black!55] at (2.3, 1.5) {$\Omega$};

    \draw[dash dot, thick, red!70!black] (0, -0.45) -- (0, 3.45);
    \node[anchor=north, font=\scriptsize\bfseries, text=red!70!black]
        at (0, -0.48) {Symmetry line ($x=0$)};

    \draw[thick, black!65] (-0.24, 0.35) -- (-0.24, 2.65);
    \fill[pattern=north east lines, pattern color=black!40]
        (-0.44, 0.35) rectangle (-0.24, 2.65);
    \foreach \y in {0.75, 1.5, 2.25} {
        \draw[thick, black!75] (-0.12, \y) circle (0.12);
    }
    \node[anchor=east, font=\scriptsize, text=black!85]
        at (-0.44, 1.5) {$u_x = 0$};

    \draw[thick, black!75] (4.6, -0.12) circle (0.12);
    \draw[thick, black!65] (4.25, -0.24) -- (4.95, -0.24);
    \fill[pattern=north east lines, pattern color=black!40]
        (4.25, -0.40) rectangle (4.95, -0.24);
    \node[anchor=north, font=\scriptsize, text=black!85]
        at (4.6, -0.42) {$u_y = 0$};

    \draw[ultra thick, -{Stealth[length=3.5mm, width=2.2mm]}, red!80!black]
        (0, 3.85) -- (0, 3.0);
    \node[anchor=west, font=\scriptsize\bfseries, text=red!80!black, xshift=3pt]
        at (0.1, 3.45) {$\vect{F}$};

    \draw[<->, thin, black!70] (0, -0.25) -- (4.1, -0.25)
        node[midway, fill=white, inner sep=1.5pt, font=\scriptsize] {$L$};

    \draw[<->, thin, black!70] (4.95, 0) -- (4.95, 3.0)
        node[midway, fill=white, inner sep=1.5pt, font=\scriptsize, rotate=90]
        {$H$};

    \draw[thick, dashed, orange!90!black] (3.4, 1.8) rectangle (4.4, 2.8);
    \draw[thick, dashed, orange!85!black, ->]
        (3.9, 2.8) to[out=50, in=145] (6.0, 2.7);
\end{scope}

\begin{scope}[shift={(6.2, 0)}]
    \node[anchor=south west, font=\small\bfseries] at (0.2, 3.65)
        {(b) Micro dual discretization};

    \fill[black!26] (0, 1.485) -- (0, 3.0) -- (1.515, 3.0) -- cycle;
    \fill[black!10] (0, 0) -- (0, 1.485) -- (1.515, 3.0) -- (3.0, 3.0)
                    -- (3.0, 1.515) -- (1.485, 0) -- cycle;

    \foreach \i in {0, 1, 2, 3} {
        \draw[black!30, thin] (\i, 0) -- (\i, 3.0);
        \draw[black!30, thin] (0, \i) -- (3.0, \i);
    }

    \draw[<->, thin, black!65] (0, -0.2) -- (1.0, -0.2)
        node[midway, below, font=\tiny] {$\delta x$};

    \draw[thick, orange!90!black, fill=orange!15, fill opacity=0.3]
        (1.0, 1.0) rectangle (2.0, 2.0);

    \node[draw=orange!90!black, fill=orange!85!black, shape=circle,
          inner sep=1.8pt] (EC) at (1.5, 1.5) {};
    \node[anchor=north, font=\tiny\bfseries, text=orange!90!black, yshift=-2pt]
        at (1.5, 1.45) {$\bar{\phi}_e$};

    \coordinate (N1) at (1.0, 1.0); 
    \coordinate (N2) at (2.0, 1.0); 
    \coordinate (N3) at (2.0, 2.0); 
    \coordinate (N4) at (1.0, 2.0); 

    \draw[->, thick, orange!80!black, opacity=0.85] (N1) -- (EC);
    \draw[->, thick, orange!80!black, opacity=0.85] (N2) -- (EC);
    \draw[->, thick, orange!80!black, opacity=0.85] (N3) -- (EC);
    \draw[->, thick, orange!80!black, opacity=0.85] (N4) -- (EC);

    \foreach \ix in {0, 1, 2, 3} {
        \foreach \iy in {0, 1, 2, 3} {
            \fill[blue!70!black] (\ix, \iy) circle (1.8pt);
        }
    }

    \node[anchor=south east, font=\tiny, text=blue!85!black]
        at (0.98, 1.02) {$\phi_1$};
    \node[anchor=north west, font=\tiny, text=blue!85!black]
        at (2.02, 0.98) {$\phi_2$};
    \node[anchor=south west, font=\tiny, text=blue!85!black]
        at (2.02, 2.02) {$\phi_3$};
    \node[anchor=south east, font=\tiny, text=blue!85!black]
        at (0.98, 2.02) {$\phi_4$};

    \draw[<->, thick, blue!85!black] (0.015, 1.50) -- (1.50, 0.015)
        node[midway, fill=white, fill opacity=0.92, inner sep=1.2pt,
             text opacity=1, font=\tiny\bfseries, text=blue!90!black,
             rotate=-45] {$w$};

    \node[anchor=north west, fill=white, fill opacity=0.95, draw=black!25,
          rounded corners=2pt, text opacity=1, inner sep=3.5pt]
        at (3.2, 3.15) {
        \scriptsize
        \begin{tabular}{@{}l@{}}
        \textbf{Duality \& Projection:}\\[3pt]
        \tikz\fill[blue!70!black] (0,0) circle (1.8pt);~%
            \textbf{Nodes} $\Nn = (n_x+1)(n_y+1)$\\
        ~~Field variable: $\phi_i \in [0, 1]$\\[3pt]
        \tikz\fill[orange!85!black] (0,0) circle (1.8pt);~%
            \textbf{Elements} $\Ne = n_x \times n_y$\\
        ~~Elastic density: $\bar{\phi}_e \in [0, 1]$\\[3pt]
        $\bar{\phi}_e = \frac{1}{4}\sum_{k=1}^4 \phi_k$\\[2pt]
        \texttt{conv2(phi,[1 1;1 1]/4,'valid')}\\[5pt]
        \textbf{Diffuse Interface:}\\[2pt]
        Width: $w$ ($2 \sim 3$ elem.)
        \end{tabular}
    };
\end{scope}

\end{tikzpicture}}
\caption{Baseline problem configuration and dual-grid spatial layout in
\texttt{topAC99}: (a)~macro half-MBB domain $\Omega$ with symmetry/roller
supports and point load $\vect{F}$; (b)~staggered nodal design variables $\phi_i$
and element physical densities $\bar{\phi}_e$ coupled matrix-free via $2\times 2$
\texttt{conv2} kernels.}\label{fig:domain_schematic}
\end{figure*}

To avoid index reshaping inside nonlinear operations, all nodal fields
(\texttt{phi}, \texttt{M}, \texttt{h}, \texttt{lambda\_h}, \texttt{grad\_c},
\texttt{grad\_h}) are stored directly as $(n_y+1)\times(n_x+1)$ 2D matrices in
MATLAB's column-major layout (rows $1\dots n_y+1$ running top-to-bottom, columns
$1\dots n_x+1$ running left-to-right), while element arrays (\texttt{x},
\texttt{ce}, \texttt{dc}) have dimensions $n_y\times n_x$. Figure~\ref{fig:domain_schematic}
illustrates the baseline half-MBB boundary conditions alongside this staggered
nodal--element grid pairing, and Table~\ref{tab:symbol_mapping} maps the
mathematical operators to source-code variables and line numbers.

\begin{table*}[t]
\centering
\footnotesize
\caption{Mapping between mathematical quantities and MATLAB arrays in
\texttt{topAC99.m}.}\label{tab:symbol_mapping}
\setlength{\tabcolsep}{2.8pt}
\begin{tabular*}{\textwidth}{@{\extracolsep{\fill}}lllll@{\extracolsep{\fill}}}
\toprule
Symbol & Variable & Dimensions & Lines & Computational Role \\
\midrule
$\ell_{\mathrm{AC}},\, \epsilon$ & \texttt{ell}, \texttt{ep} & scalar, scalar & 3--4 & Correlation length and gradient parameter (avoids \texttt{eps}) \\
$\vect{\phi},\, \bar{\vect{\phi}}$ & \texttt{phi}, \texttt{x} & $(n_y+1)\times(n_x+1)$, $n_y\times n_x$ & 30, 35, 91 & Nodal design variables and element physical densities \\
$\mat{K}_{\mathrm{lap}}$ & \texttt{Klap} & $\Nn\times\Nn$ (5-pt sparse) & 28 & Lumped tensor-product Laplacian ($\vect{1}^\top\mat{K}_{\mathrm{lap}} = \vect{0}^\top$) \\
$\mat{M},\, \mat{M}^{-1}$ & \texttt{M}, \texttt{Minv} & $(n_y+1)\times(n_x+1)$ & 27 & Trapezoidal nodal tributary weights and pointwise inverse \\
$\vect{h}(\vect{\phi}),\, \|\vect{h}\|_*$ & \texttt{h}, \texttt{h\_norm} & $(n_y+1)\times(n_x+1)$, scalar & 48--49 & Conserved Allen--Cahn nodal residual and dual RMS norm \\
$\vect{\lambda}_h,\, \rho_h$ & \texttt{lambda\_h}, \texttt{rho\_h} & $(n_y+1)\times(n_x+1)$, scalar & 32, 93--96 & Nodal ALM multipliers (zero-mean projected) and penalty scalar \\
$\nabla_{\vect{\phi}}c,\, \mat{H}^\top\vect{z}$ & \texttt{grad\_c}, \texttt{grad\_h} & $(n_y+1)\times(n_x+1)$ & 43, 52 & Adjoint objective sensitivity and matrix-free AC Jacobian action \\
$L_i, U_i, a_i, b_i, \lambda$ & \texttt{low}, \texttt{upp}, \texttt{a\_lim}, \texttt{b\_lim}, \texttt{lmid} & $(n_y+1)\times(n_x+1)$, scalar & 31, 62--72, 77 & Moving asymptotes, trust-region bounds, and volume dual multiplier \\
\bottomrule
\end{tabular*}
\end{table*}

\subsection{Modular Software Walkthrough (Lines 1--99)}\label{subsec:walkthrough}

The script is partitioned into five self-contained functional blocks, with default
algorithmic parameters summarized in Table~\ref{tab:param_summary}:

\begin{table*}[t]
\centering
\footnotesize
\caption{Summary of physical and algorithmic parameters in \texttt{topAC99.m}.}\label{tab:param_summary}
\setlength{\tabcolsep}{5pt}
\begin{tabular*}{\textwidth}{@{\extracolsep{\fill}}lcclp{9.2cm}@{\extracolsep{\fill}}}
\toprule
Parameter & Code & Default & Typical Range & Physical and Algorithmic Function \\
\midrule
$w$ & \texttt{w} & $2.1$ & $2.0$--$4.5$ & Diffuse $0.1\to 0.9$ interface width (elements); controls member transition and suppresses 5-pt axial bias for $w \ge 2.0$. \\
$\beta$ & \texttt{beta} & $0.065$ & $0.001$--$0.12$ & Double-well barrier scale; order-of-magnitude initial estimate via Eq.~\eqref{eq:beta_scaling} ($0.065$ for compliance; $0.002$ for force inverter). \\
$p$ & \texttt{penal} & $3.0$ & $3.0$--$4.0$ & SIMP stiffness interpolation exponent. \\
$\rho_h$ & \texttt{rho\_h} & $0.10$ & $0.05$--$0.50$ & Initial ALM quadratic penalty; multiplied by $1.1$ ($\le 10$) every 10 loops when $\|\vect{h}\|_* > 0.015\beta$. \\
$\delta$ & \texttt{move} & $0.10$ & $0.02$--$0.20$ & Initial MMA trust-region step bound ($\delta < 0.5$); annealed by $\times 0.95$ ($\ge 0.005$) once $|c - c_{\mathrm{old}}|/\max(|c|,10^{-6}) < 10^{-3}$ and $\|\vect{h}\|_* \le 0.28\beta$ after loop 30. \\
$w_h$ & \texttt{w\_h} & ramp & $0.11\to 1.0$ & Continuation weight ramping linearly over 16 iterations to prevent premature interface locking. \\
\bottomrule
\end{tabular*}
\end{table*}

\begin{itemize}
    \item \textbf{Module 0--1: Interface Conversion \& FEA Setup (Lines 1--20).}
        Lines 3--4 convert $(w,\beta)$ into $(\ell_{\mathrm{AC}}, \epsilon)$ via
        Eq.~\eqref{eq:param_conversion}, naming $\epsilon$ as \texttt{ep} so
        MATLAB's built-in floating-point accuracy function \texttt{eps} is never
        shadowed. Lines 6--16 retain the concise 64-entry Q4 triplet indexing
        (\texttt{edofMat}, \texttt{iK}, \texttt{jK}) of \texttt{top88}~\cite{Andreassen2011}
        to stay within 99 lines without external \texttt{fsparse} MEX routines
        (see Section~\ref{sec:timing_breakdown} for memory trade-offs vs.\ 36-entry
        half-stiffness assembly~\cite{Ferrari2020b}). Lines 17--20 isolate state
        \texttt{U}, load \texttt{F}, and restraints \texttt{fixeddofs}/\texttt{freedofs}.
    \item \textbf{Module 2: Kronecker-Sum Laplacian \& Mass Arrays (Lines 21--28).}
        Lines 22--25 construct 1D Neumann diffusion matrices \texttt{Dx} and
        \texttt{Dy} via \texttt{spdiags} with corner diagonal entries set to $1$.
        Lines 26--28 form trapezoidal mass arrays \texttt{M}, \texttt{Minv} and
        assemble the 2D Kronecker Laplacian \texttt{Klap} in a single expression
        (Eq.~\eqref{eq:kronecker_lap}).
    \item \textbf{Module 3: State \& ALM Initialization (Lines 29--32).}
        Initializes the homogeneous nodal field $\texttt{phi} = V_0$, two-step
        history buffers \texttt{p\_old1}, \texttt{p\_old2}, and initial
        asymptotes $\texttt{low} = \texttt{phi} - 0.5$, $\texttt{upp} =
        \texttt{phi} + 0.5$ (requiring $\delta < 0.5$ so $\vect{\phi}^{(2)}$
        stays inside $(\vect{L}, \vect{U})$ at loop 2). It also sets zero
        multipliers \texttt{lambda\_h}, penalty $\texttt{rho\_h} = 0.1$, and
        trust radius $\texttt{move} = 0.10$.
    \item \textbf{Module 4: Primal/Adjoint Mechanics, AC-ALM \& 1D-Dual MMA
        (Lines 33--85).}
        Inside the main loop, Line 35 maps $\texttt{phi}\to\texttt{x}$ via
        \texttt{'valid'} convolution. Lines 37--43 assemble and symmetrize the
        global stiffness \texttt{K} (\texttt{(K+K')/2} for Cholesky dispatch in
        \texttt{mldivide}), solve $\mat{K}\vect{U}=\vect{F}$, evaluate compliance
        and element sensitivities \texttt{dc}, and backpropagate \texttt{grad\_c}
        via \texttt{'full'} convolution. Lines 45--56 evaluate $W'$, $W''$,
        $\mu$, residual $\vect{h}$, dual norm $\|\vect{h}\|_*$, and matrix-free
        Jacobian action \texttt{grad\_h} (Eq.~\eqref{eq:fast_Hzeta}), forming
        total sensitivity $\texttt{df0} = \texttt{grad\_c} + w_h\,\texttt{grad\_h}$.
        Line 57 implements zero-crossing-safe, sign-agnostic trust-region annealing:
        \texttt{abs(c - c\_old)/max(abs(c),1e-6) < 1e-3 \&\& h\_norm <= 0.28*beta}.
        Physically, the spinodal barrier maximum of $|W'(\phi)| = \beta|2\phi -
        6\phi^2 + 4\phi^3|$ on $[0,1]$ occurs at $\phi = \frac{1}{2} \pm
        \frac{1}{2\sqrt{3}}$, where $|W'(\phi)|_{\max} = \frac{\beta}{3\sqrt{3}}
        \approx 0.1925\beta$. The gate $\|\vect{h}\|_* \le 0.28\beta \approx
        1.45|W'|_{\max}$ therefore requires the domain-RMS Allen--Cahn residual
        to subside to the intrinsic spinodal restoring scale before trust-region
        contraction begins. Lines 58--65 adapt moving asymptotes ($\times 1.2$
        for monotone steps, $\times 0.7$ for sign reversals), and Lines 67--85
        execute the 18-line 1D-dual bisection MMA update
        (Eq.~\eqref{eq:mma_primal_sol}).
    \item \textbf{Module 5: Monitoring, Dual Stopping Gate \& Multiplier Update
        (Lines 86--99).}
        Lines 87--89 log iteration telemetry and render physical densities (using
        \texttt{caxis([0 1])} for backward and GNU Octave 7.x compatibility). Line 90 enforces a dual
        stopping gate ($\text{loop} \ge 30$, $\text{ch} < 0.01$, and
        $\|\vect{h}\|_* \le 0.28\beta$) prior to state update so returned quantities
        $(c,\texttt{x},\text{loop},\|\vect{h}\|_*)$ correspond strictly to the
        evaluated iterate. Lines 91--97 commit \texttt{phi\_trial} and perform
        clamped, zero-mean-projected multiplier updates alongside scale-consistent
        penalty adaptation ($\|\vect{h}\|_* > 0.015\beta$).
\end{itemize}

\subsection{Architectural Generalization to Three Dimensions}\label{subsec:3d_arch}

Although \texttt{topAC99.m} targets 2D Q4 grids to fit within 99 lines, every
Allen--Cahn and optimization module generalizes directly to 3D structured
hexahedral (H8) grids of $(n_x+1)\times(n_y+1)\times(n_z+1)$ nodes without
structural redesign. First, the 5-point Kronecker Laplacian
(Eq.~\eqref{eq:kronecker_lap}) extends to a 7-point three-way Kronecker sum
$\mat{D}_x \otimes \mat{M}_y \otimes \mat{M}_z + \mat{M}_x \otimes \mat{D}_y
\otimes \mat{M}_z + \mat{M}_x \otimes \mat{M}_y \otimes \mat{D}_z$ with
width-invariant $\mathcal{O}(\Nn)$ storage. Second, dual-grid primal averaging
and adjoint backpropagation (Eq.~\eqref{eq:conv2_duality}) replace
\texttt{conv2} with native 3D convolution \texttt{convn(..., ones(2,2,2)/8,
'valid'/'full')}. Third, pointwise evaluations of $W', W''$ and the 18-line
1D-dual bisection MMA solver are completely dimension-independent.
Consequently, just as in 3D SIMP frameworks~\cite{LiuKai2014a,Ferrari2020b}, the
sole computational bottleneck in 3D lies in replacing direct sparse Cholesky
factorization (\texttt{K\textbackslash F}) with matrix-free preconditioned
conjugate gradients or geometric multigrid.

\section{Verification, Computational Scaling, and Parameter Physics}\label{sec:benchmarks}

All numerical verifications and timings were executed in single-thread mode
(\texttt{maxNumCompThreads(1)}) on a Debian 12 Linux workstation equipped with an
Intel Core i7-11850H processor (2.50\,GHz) and MATLAB R2023a, with console and
graphics I/O (Lines 87--89) suppressed equally across all compared codes during
headless timing runs.

\subsection{Baseline Verification and Multi-Resolution Scaling}\label{subsec:mbb_benchmark}

To separate intrinsic topological stiffness from diffuse-interface SIMP
penalization across methods with differing boundary sharpness, we report
Sigmund's continuous measure of non-discreteness~\cite{Sigmund2007}:
\[
M_{\mathrm{nd}} = \frac{4}{\Ne}\sum_{e=1}^{\Ne}\bar{\phi}_e(1 - \bar{\phi}_e) \times 100\%.
\]
We also evaluate a volume-preserving binary compliance $c_{0/1}$ by
thresholding the highest-density elements up to the target volume fraction
$V_0$ to unity and setting the remainder to zero.

\begin{figure}[htbp]
\centering
\begin{subfigure}[b]{0.48\columnwidth}
    \centering
    \includegraphics[width=\textwidth]{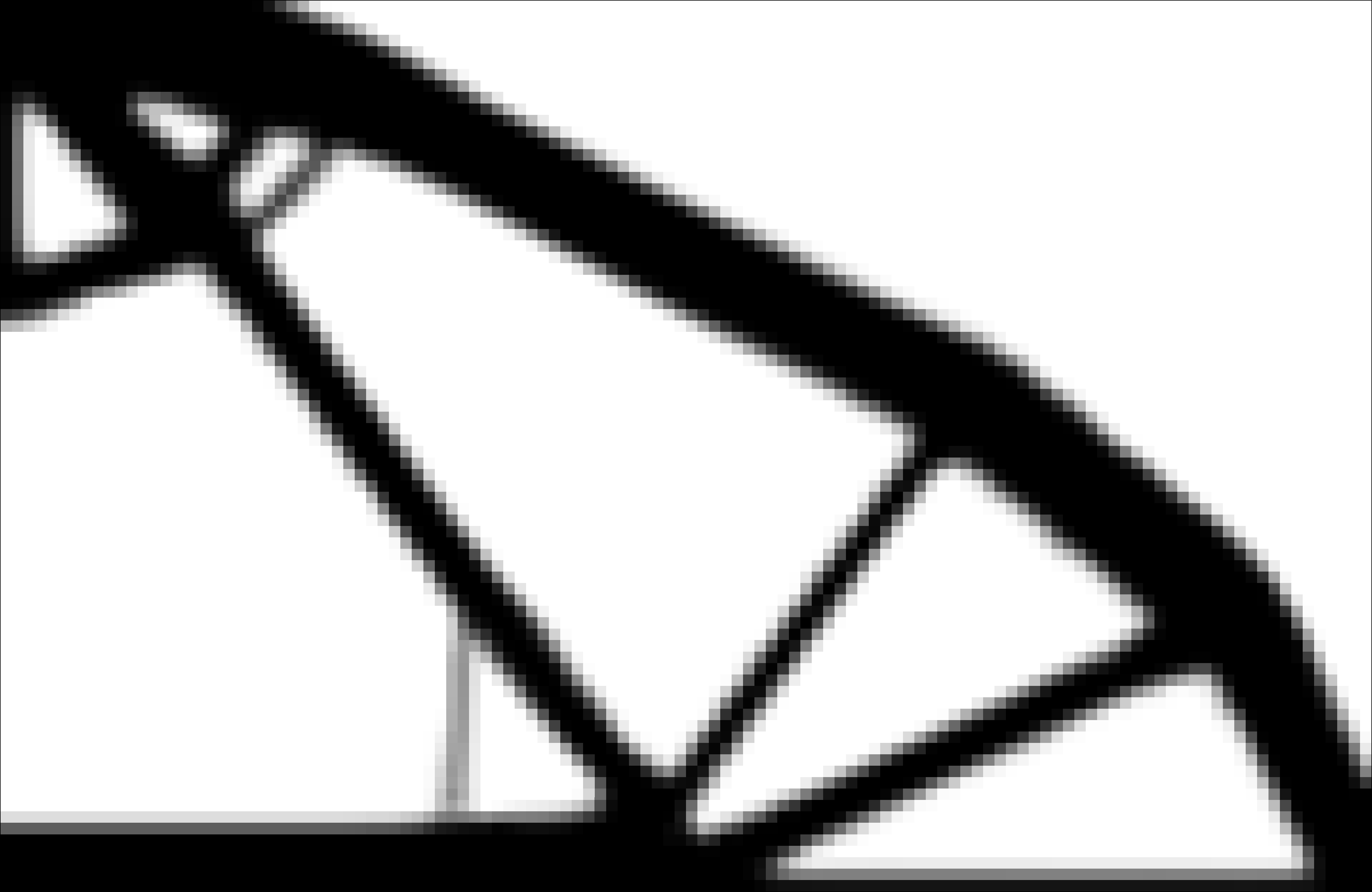}
    \caption{$V_0 = 0.30$ ($c = 77.435$, 80 iters)}
    \label{fig:mbb_v03}
\end{subfigure}
\hfill
\begin{subfigure}[b]{0.48\columnwidth}
    \centering
    \includegraphics[width=\textwidth]{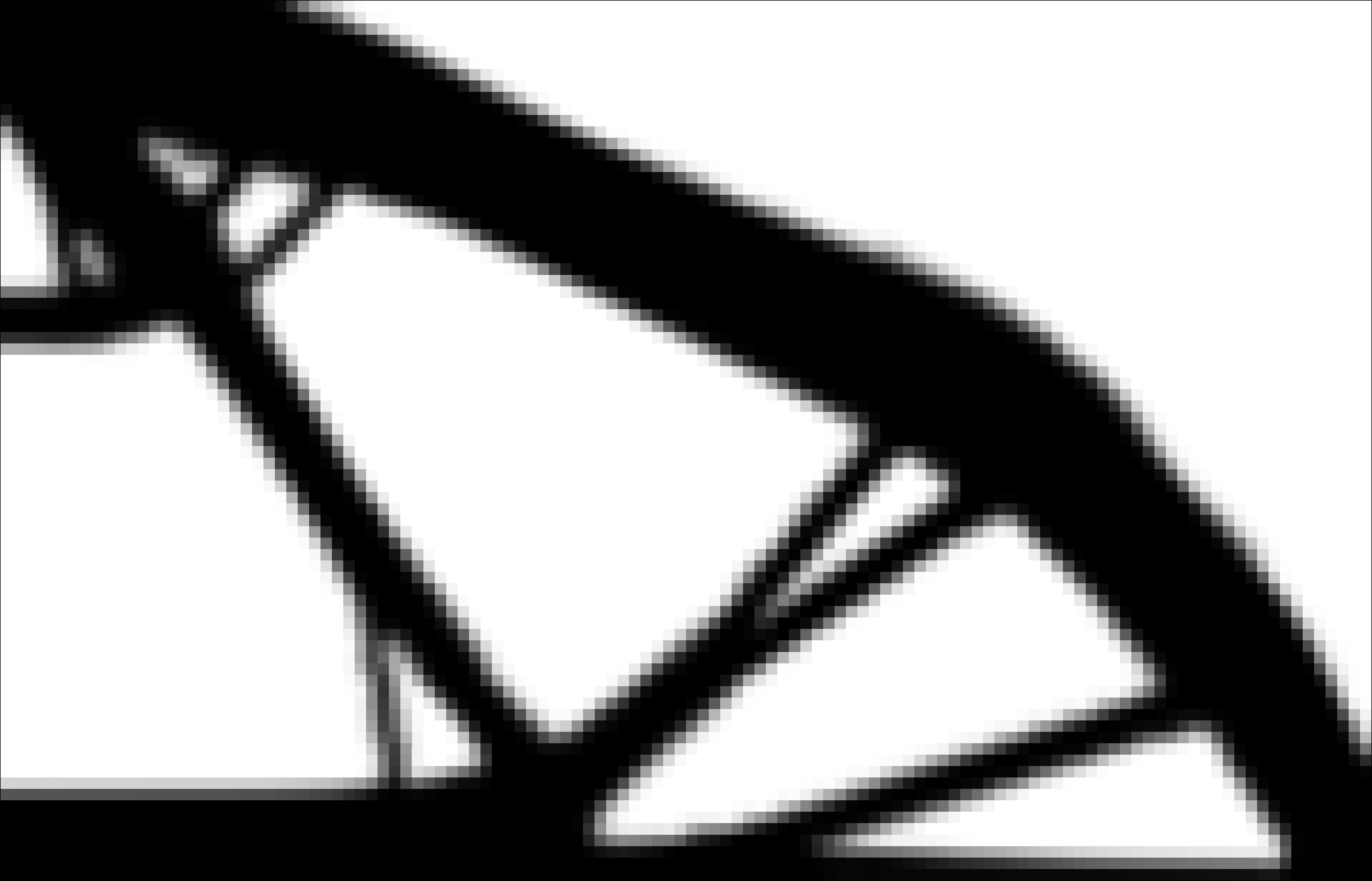}
    \caption{$V_0 = 0.40$ ($c = 59.163$, 81 iters)}
    \label{fig:mbb_v04}
\end{subfigure}
\caption{Converged half-MBB topologies on a $120\times 80$ grid ($w = 2.1$,
$\beta = 0.065$).}\label{fig:mbb_benchmark}
\end{figure}

\begin{figure}[htbp]
\centering
\includegraphics[width=\columnwidth]{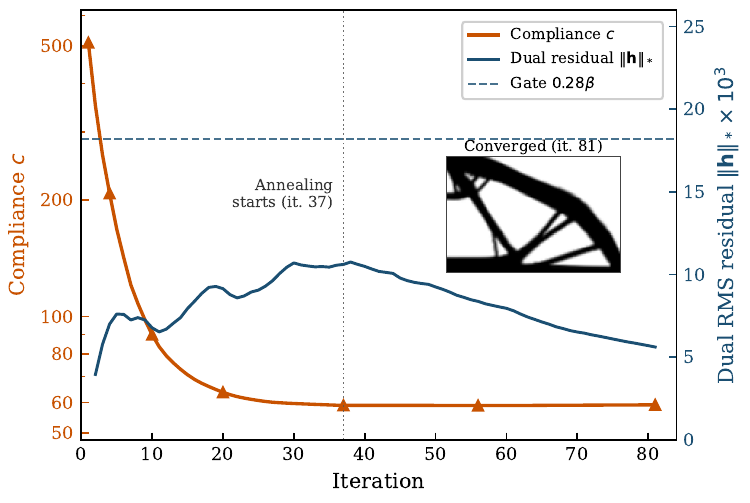}
\caption{Dual convergence trajectory of compliance $c$ (left semi-log axis) and
Allen--Cahn RMS residual $\|\vect{h}\|_*$ (right axis, with $0.28\beta$ annealing
threshold) alongside the converged topology inset on the $120\times 80$
half-MBB beam ($V_0 = 0.40$).}\label{fig:convergence_history}
\end{figure}

We first verify \texttt{topAC99} on the half-MBB beam using baseline settings
($p = 3.0$, $w = 2.1$, $\beta = 0.065$). Figure~\ref{fig:mbb_benchmark} shows
cleanly converged $120\times 80$ topologies for $V_0 = 0.30$ and $V_0 = 0.40$,
while Fig.~\ref{fig:convergence_history} tracks the dual history of compliance
and Allen--Cahn residual for $V_0 = 0.40$. Starting from a uniform field, the
primary load-transfer trusses nucleate within the first 25 iterations at the
initial trust radius. By iteration~37, both relative objective stabilization
and spinodal stationarity ($\|\vect{h}\|_* \le 0.28\beta$) are satisfied,
engaging exponential move-limit annealing ($\times 0.95$ per step) that smoothly
sharpens phase boundaries until convergence at iteration~81 with tight residual
and volume accuracy (Table~\ref{tab:benchmark_comparison}).

Table~\ref{tab:benchmark_comparison} evaluates solver scaling across a 25-fold
increase in mesh resolution from $60\times 40$ to $300\times 200$ at fixed
stencil thickness $w = 2.1$ elements:
\begin{enumerate}
    \item \textit{Comparison with full-FEM AC-ALM~\cite{DuRun2026topac} and
        role of move-limit annealing}: On $120\times 80$ ($V_0 = 0.30$),
        \texttt{topAC99} reaches $c = 77.435$ in 80 iterations ($3.59$\,s),
        compared with $c = 80.75$ in 363 iterations for the consistent
        9-point/$3\times 3$-quadrature code~\cite{DuRun2026topac}. This
        $4.5\times$ iteration reduction reflects a deliberate engineering
        trade-off: relaxing the normalized residual gate from $0.075$ to
        $0.28$ and engaging $\times 0.95$ trust-region contraction drives the
        step bound from $0.10$ below $0.01$ within 45 annealing steps. Across
        refined grids ($120\times 80$ to $300\times 200$), outer iteration counts
        remain nearly resolution-independent at roughly $80$--$90$ steps, while
        total runtime is governed primarily by the sparse Cholesky factorization
        of the elastic system (examined further in Section~\ref{sec:timing_breakdown}).
    \item \textit{Disentangling FEA softening from diffuse-layer grayscale}:
        Holding $w$ fixed in \emph{element} units thins the transition layer
        relative to the domain size as the mesh is refined, driving both
        $M_{\mathrm{nd}}$ and the Allen--Cahn residual $\|\vect{h}\|_*$ down
        monotonically while keeping volume errors negligible. At $V_0 = 0.30$,
        raw compliance $c$ decreases with refinement because reduced boundary
        grayscale outweighs element stiffening; once binary thresholding removes
        diffuse-layer SIMP penalization, however, $c_{0/1}$ increases
        monotonically with mesh resolution at both volume fractions, matching
        classical Q4 displacement finite-element convergence from below.
\end{enumerate}

\begin{table*}[t]
\centering
\footnotesize
\caption{Multi-resolution scaling performance on the half-MBB benchmark at fixed
element interface width ($p = 3.0$, $w = 2.1$, $\beta = 0.065$, single-thread
headless MATLAB R2023a).}\label{tab:benchmark_comparison}
\setlength{\tabcolsep}{5.0pt}
\begin{tabular*}{\textwidth}{@{\extracolsep{\fill}}lrrccrrr@{\extracolsep{\fill}}}
\toprule
Grid ($n_x \times n_y$) & Comp.\ $c$ & $c_{0/1}$ & Iter. & $M_{\mathrm{nd}}$ (\%) & $\|\vect{h}\|_*$ & $|V - V_0|$ & Time (s) \\
\midrule
\multicolumn{8}{l}{\textit{Target volume fraction $V_0 = 0.30$}} \\
$60 \times 40$   & $81.551$ & $69.865$ & $116$ & $17.65\%$ & $1.46 \times 10^{-2}$ & $2.01 \times 10^{-6}$ & $1.16$ \\
$120 \times 80$  & $77.435$ & $70.616$ & $80$  & $11.16\%$ & $6.92 \times 10^{-3}$ & $3.09 \times 10^{-6}$ & $3.59$ \\
$180 \times 120$ & $76.590$ & $71.197$ & $89$  & $8.91\%$  & $3.93 \times 10^{-3}$ & $4.30 \times 10^{-6}$ & $10.14$ \\
$240 \times 160$ & $76.234$ & $71.902$ & $92$  & $7.40\%$  & $2.50 \times 10^{-3}$ & $4.14 \times 10^{-6}$ & $21.75$ \\
$300 \times 200$ & $76.376$ & $72.480$ & $92$  & $6.47\%$  & $1.77 \times 10^{-3}$ & $3.46 \times 10^{-6}$ & $33.98$ \\
\midrule
\multicolumn{8}{l}{\textit{Target volume fraction $V_0 = 0.40$}} \\
$60 \times 40$   & $59.316$ & $54.465$ & $81$  & $15.79\%$ & $1.21 \times 10^{-2}$ & $2.68 \times 10^{-6}$ & $0.88$ \\
$120 \times 80$  & $59.163$ & $55.416$ & $81$  & $11.64\%$ & $5.62 \times 10^{-3}$ & $5.10 \times 10^{-6}$ & $3.75$ \\
$180 \times 120$ & $59.613$ & $56.414$ & $78$  & $10.36\%$ & $3.34 \times 10^{-3}$ & $9.67 \times 10^{-6}$ & $9.01$ \\
$240 \times 160$ & $59.773$ & $57.186$ & $83$  & $8.40\%$  & $2.16 \times 10^{-3}$ & $7.89 \times 10^{-6}$ & $19.26$ \\
$300 \times 200$ & $60.415$ & $58.058$ & $86$  & $7.91\%$  & $2.16 \times 10^{-3}$ & $4.17 \times 10^{-6}$ & $34.30$ \\
\bottomrule
\end{tabular*}
\end{table*}

\subsection{Modular Execution Profiling and Horizontal Benchmark Comparison}%
\label{sec:timing_breakdown}

\begin{figure}[htbp]
\centering
\includegraphics[width=\columnwidth]{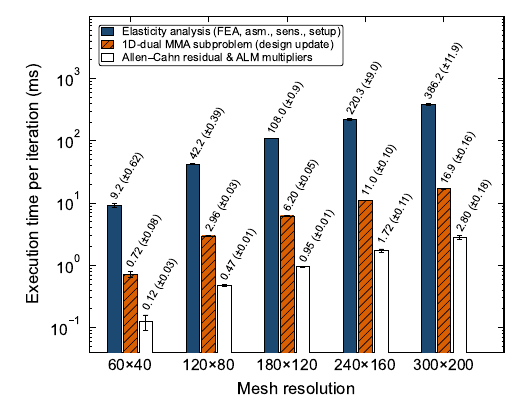}
\caption{Per-iteration modular execution time breakdown (mean $\pm$ standard
deviation across 5 headless runs) on the half-MBB beam ($V_0 = 0.30$).}\label{fig:timing_breakdown}
\end{figure}

Figure~\ref{fig:timing_breakdown} breaks down the per-iteration headless execution
time across all five mesh resolutions ($V_0 = 0.30$, averaged over five
independent runs). Forward elasticity analysis (stiffness assembly, Cholesky
factorization, compliance sensitivity, and amortized setup) dominates execution,
absorbing $91.5\%$ of total runtime on $60\times 40$ and rising to $95.1\%$ on
$300\times 200$. By contrast, the vectorized 1D-dual MMA bisection solver
consumes only $4.2\%$--$7.2\%$ of step time, while the lumped Allen--Cahn
residual, matrix-free adjoint action, and ALM multiplier updates require merely
$0.7\%$--$1.2\%$ (and one-off Kronecker setup on $120\times 80$ takes just
$16.4$\,ms). Importantly, this $<1.2\%$ share means only that replacing spatial
density filters with variational AC-ALM regularization adds negligible
incremental cost per iteration---not that an individual iteration runs faster
than in SIMP, since per-step runtime remains governed by the forward elastic
solve.

Table~\ref{tab:horizontal_comparison} compares \texttt{topAC99} on
$120\times 80$ and $300\times 200$ grids against three compact MATLAB
baselines at a variance-matched~\cite{Lazarov2011a} physical length scale
($\ell_{\mathrm{AC}} = \ell_{\mathrm{PDE}} = 0.68$, equivalent to conical
filter radius $r_{\min} = 2.34$). The compared baselines are:
(i)~sparse-matrix density-filtered \texttt{top88}~\cite{Andreassen2011};
(ii)~Helmholtz PDE-filtered SIMP~\cite{Lazarov2011a} (\texttt{top82}~\cite{Andreassen2011}
with AMD-permuted Cholesky pre-factorization); and (iii)~a toolbox-free
adaptation of \texttt{top99neo}~\cite{Ferrari2020b} (half-stiffness assembly
and \texttt{conv2} filtering) using both linear filtering and 250-step
Heaviside $\tanh$ projection continuation.

Three architectural conclusions emerge from Table~\ref{tab:horizontal_comparison}:
\begin{enumerate}
    \item \textit{Per-step cost vs.\ converged runtime and termination-rule effects}:
        On a per-iteration basis, \texttt{topAC99} ($44.9$\,ms on $120\times 80$;
        $369$\,ms on $300\times 200$) is governed by elastic FEA. On $120\times 80$
        it is slightly slower per step than \texttt{top99neo}-base ($36.3$\,ms,
        due to symmetric half-stiffness assembly), whereas on $300\times 200$
        both achieve nearly identical step times ($\sim 0.37$\,s) once Cholesky
        factorization dominates. By contrast, \texttt{top82} is markedly slower
        per step ($79.9$--$946$\,ms) because its OC loop executes triangular PDE
        back-solves inside every bisection trial. Under the shared termination
        criterion ($\text{ch} < 0.01$), \texttt{topAC99} reaches the lowest
        binary compliance on both grids and converges in roughly $4\times$--$5\times$
        fewer iterations than linear \texttt{top99neo}-base, reflecting MMA
        convergence with move-limit annealing once $\|\vect{h}\|_* \le 0.28\beta$.
        On $300\times 200$, \texttt{top88} and \texttt{top82} instead exhaust the
        500-iteration ceiling due to OC boundary chattering, while
        \texttt{top99neo}-base with projection (\texttt{ft=3}) executes a fixed
        250-step continuation schedule; their wall-clock times are therefore
        listed for completeness only.
    \item \textit{Diffuse equilibrium vs.\ Heaviside projection sharpness}:
        Heaviside projection produces the sharpest raw designs, whereas
        \texttt{topAC99} retains a smooth $\tanh$ transition layer of thickness
        $w = 2.1$. On the finer $300\times 200$ grid at small filter radius,
        however, aggressive $\beta_{\mathrm{H}}$ continuation locks the projected
        SIMP design into a visibly stiffer local minimum, whereas
        \texttt{topAC99}'s ramped weight $w_h$ allows free topological nucleation
        before spinodal sharpening.
    \item \textit{Regularizer vs.\ FEA memory footprint}: On $300\times 200$,
        \texttt{topAC99}'s persistent regularizer arrays (\texttt{Klap},
        \texttt{M}, \texttt{Minv}) require $3.5\times$ less RAM than
        \texttt{top88}'s sparse filter and $8.7\times$ less than \texttt{top82}'s
        Cholesky factor, while avoiding $4.32$\,MB of sparse node--element
        transfer storage via \texttt{conv2}. Moreover, expanding the filter radius
        from $2.1$ to $8.4$ on $300\times 200$ inflates explicit filter assembly
        buffers from $36.0$\,MB to $416.2$\,MB and sparse filter storage from
        $13.0$\,MB to $212.6$\,MB, whereas $\mat{K}_{\mathrm{lap}}$ remains fixed
        at $6.28$\,MB ($301{,}501$ nonzeros) regardless of interface width $w$.
        Overall workspace memory is instead dominated by elastic stiffness
        arrays: our concise 64-entry assembly occupies $120.0$\,MB on
        $300\times 200$, versus $87.4$\,MB for \texttt{top99neo}'s 36-entry
        half-stiffness indexing (which can be substituted directly into Module~1
        of \texttt{topAC99} in 8 additional lines).
\end{enumerate}

\begin{table*}[t]
\centering
\footnotesize
\caption{Horizontal benchmark and workspace memory comparison on the half-MBB
beam at variance-matched length scale ($V_0 = 0.30$, $\ell_{\mathrm{AC}} =
\ell_{\mathrm{PDE}} = 0.68 \leftrightarrow r_{\min} = 2.34$, single-thread
headless MATLAB R2023a).}\label{tab:horizontal_comparison}
\setlength{\tabcolsep}{4.0pt}
\begin{tabular*}{\textwidth}{@{\extracolsep{\fill}}llrrccrrr@{\extracolsep{\fill}}}
\toprule
Grid & Method & Comp.\ $c$ & $c_{0/1}$ & Iter. & $M_{\mathrm{nd}}$ (\%) & Time (s) & Reg.\ RAM (MB)$^a$ & FEA RAM (MB)$^b$ \\
\midrule
$120 \times 80$  & \texttt{topAC99} ($w=2.1,\,\beta=0.065$)             & $77.435$ & $70.616$ & $80$         & $11.16\%$ & $3.59$   & $1.17$  & $19.36$ \\
                 & \texttt{top88} (\texttt{ft=2}, $r_{\min}=2.34$)      & $80.778$ & $71.622$ & $461$        & $13.00\%$ & $23.61$  & $4.07$  & $19.41$ \\
                 & PDE-SIMP \texttt{top82} ($\ell_{\mathrm{PDE}}=0.68$) & $82.034$ & $71.666$ & $470$        & $14.17\%$ & $37.53$  & $8.74$  & $20.15$ \\
                 & \texttt{top99neo}-base (\texttt{ft=1}, $r_{\min}=2.34$) & $81.285$ & $72.152$ & $122$     & $12.84\%$ & $4.43$   & $0.15$  & $13.98$ \\
                 & \texttt{top99neo}-base (\texttt{ft=3}, $r_{\min}=2.34$) & $79.689$ & $79.087$ & $250^\S$  & $0.89\%$  & $8.73$   & $0.15$  & $13.98$ \\
\midrule
$300 \times 200$ & \texttt{topAC99} ($w=2.1,\,\beta=0.065$)             & $76.376$ & $72.480$ & $92$         & $6.47\%$  & $33.98$  & $7.24$  & $119.97$ \\
                 & \texttt{top88} (\texttt{ft=2}, $r_{\min}=2.34$)      & $79.161$ & $75.482$ & $500^\dagger$& $5.86\%$  & $229.20$ & $25.44$ & $119.54$ \\
                 & PDE-SIMP \texttt{top82} ($\ell_{\mathrm{PDE}}=0.68$) & $79.088$ & $74.986$ & $500^\dagger$& $6.43\%$  & $473.13$ & $63.24$ & $123.40$ \\
                 & \texttt{top99neo}-base (\texttt{ft=1}, $r_{\min}=2.34$) & $79.184$ & $75.504$ & $469$     & $5.85\%$  & $175.53$ & $0.96$  & $87.37$ \\
                 & \texttt{top99neo}-base (\texttt{ft=3}, $r_{\min}=2.34$) & $86.076$ & $85.908$ & $250^\S$  & $0.44\%$  & $91.55$  & $0.96$  & $87.37$ \\
\bottomrule
\end{tabular*}
\vspace{0.3ex}
{\scriptsize $^a$Persistent regularization operators (\texttt{whos} bytes): \texttt{Klap+M+Minv} (\texttt{topAC99}); \texttt{H+Hs} (\texttt{top88}); \texttt{KF+LF+sF+TF} (\texttt{top82}); \texttt{h+Hs+dHs} (\texttt{top99neo}-base). $^b$Persistent FEA arrays (\texttt{iK/jK/Iar}, \texttt{sK}, \texttt{K}). $^\dagger$Non-terminated at 500-iteration OC cap ($\text{ch} > 0.01$; not directly comparable as converged runtime). $^\S$Prescribed 250-iteration $\beta_{\mathrm{H}}$ continuation schedule~\cite{Ferrari2020b}.}
\end{table*}

\subsection{Length-Scale Control and Order-of-Magnitude Barrier Calibration}%
\label{sec:high_res_beta}

\begin{figure}[htbp]
\centering
\begin{subfigure}[b]{0.48\columnwidth}
    \centering
    \includegraphics[width=\textwidth]{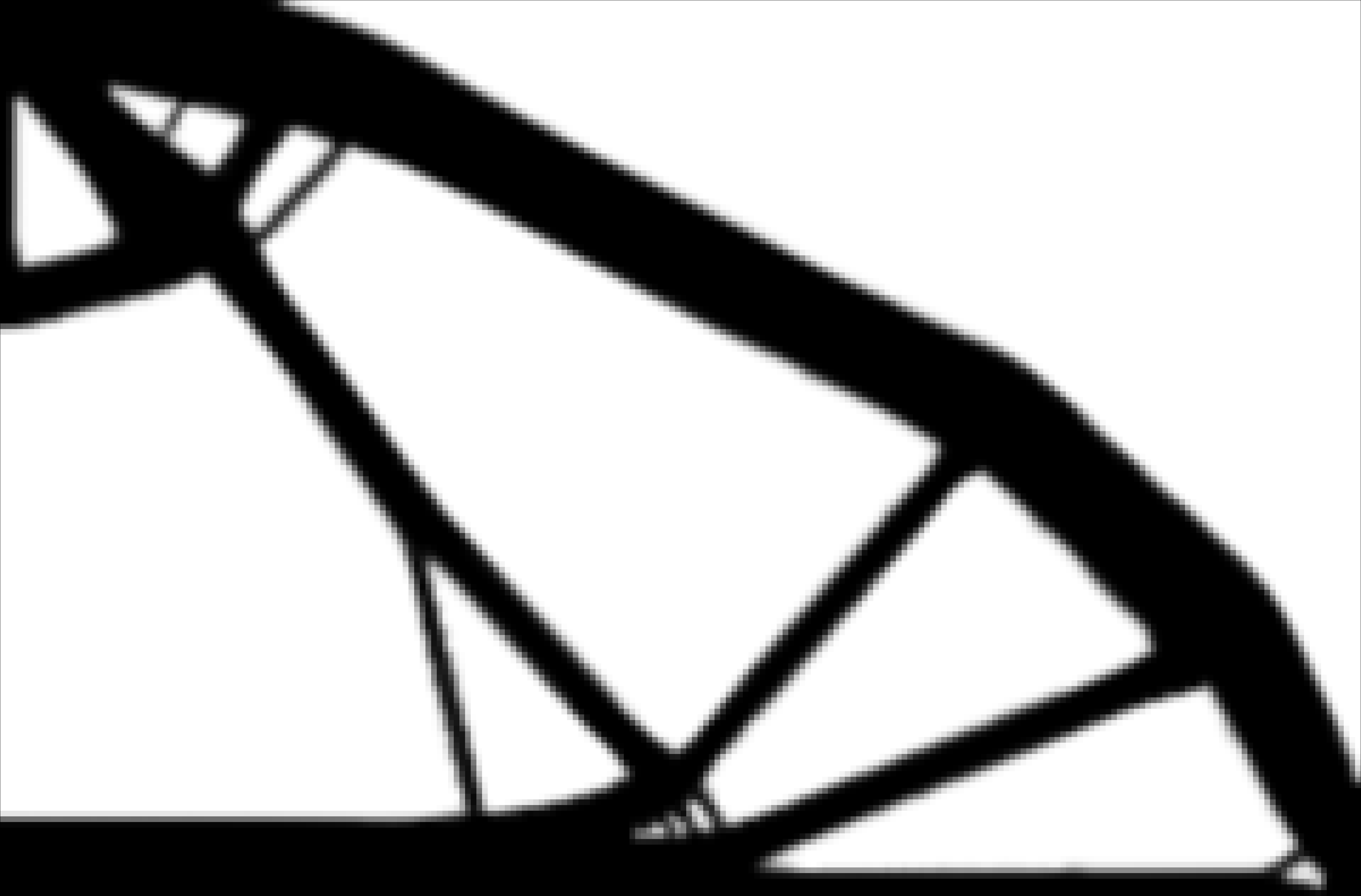}
    \caption{$w = 2.1$ ($c_{0/1} = 71.902$, 92 iters)}
    \label{fig:rect_w21}
\end{subfigure}
\hfill
\begin{subfigure}[b]{0.48\columnwidth}
    \centering
    \includegraphics[width=\textwidth]{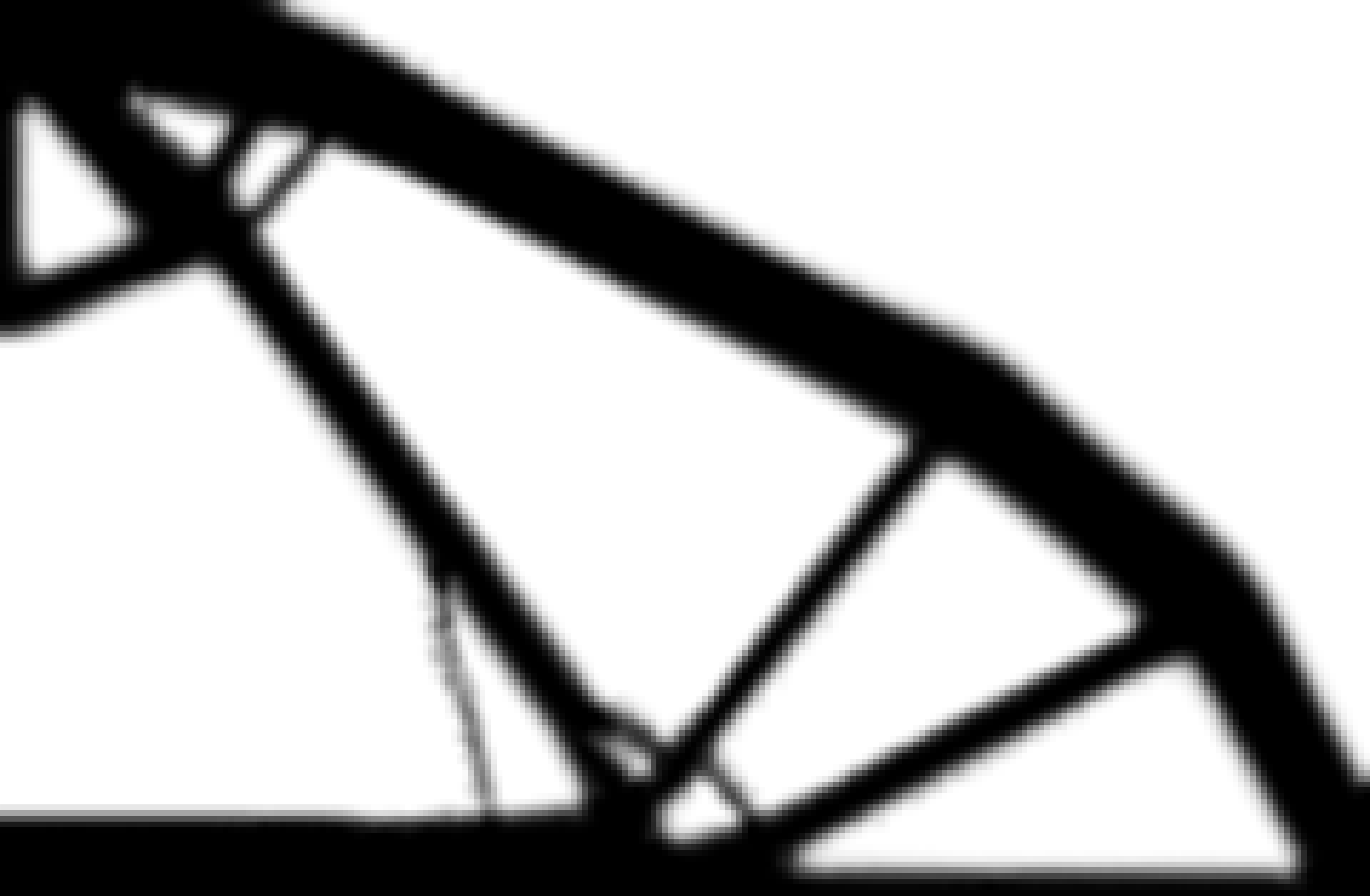}
    \caption{$w = 4.2$ ($c_{0/1} = 72.251$, 85 iters)}
    \label{fig:rect_w42}
\end{subfigure}
\caption{Length-scale regularization via interface width $w$ on a $240\times 160$
half-MBB grid ($V_0 = 0.30$, $\beta = 0.065$).}\label{fig:rect_240x160}
\end{figure}

Table~\ref{tab:high_res_comparison} and Fig.~\ref{fig:rect_240x160} examine the
decoupled roles of $(w,\beta)$ on a $240\times 160$ grid ($V_0 = 0.30$):
\begin{itemize}
    \item \textit{Interface width $w$ and physical length-scale consistency}:
        Doubling $w$ from $2.1$ to $4.2$ on $240\times 160$ suppresses thin
        secondary braces in favor of a cleaner 4-panel truss
        (Fig.~\ref{fig:rect_240x160}b; a faint partially dissolved rib in the
        lower-left bay reflects early move-limit decay). Crucially, the
        $120\times 80$ design at $w = 2.1$ (Fig.~\ref{fig:mbb_v03}) and the
        $240\times 160$ design at $w = 4.2$ (Fig.~\ref{fig:rect_w42}) share the
        exact same \emph{physical} interface ratio ($w/L_x = 0.0175$) and
        converge to closely matching macroscopic layouts and binary compliance.
        Moreover, Table~\ref{tab:high_res_comparison} confirms that the higher
        raw compliance at $w = 4.2$ arises almost entirely from wider
        diffuse-layer SIMP penalization ($M_{\mathrm{nd}}$ increases), while
        binary compliance $c_{0/1}$ varies by less than $0.5\%$.
    \item \textit{Heuristic order-of-magnitude calibration for $\beta$}:
        During early interface nucleation, when the multipliers are still near
        zero and the penalty remains at its initial value $\rho_{h,0}$,
        Eq.~\eqref{eq:param_conversion} implies that both the residual and its
        Jacobian scale linearly with $\beta$, so the initial quadratic penalty
        restoring gradient grows roughly like $w_h\rho_{h,0}\beta^2$. Balancing
        this initial penalty gradient against the mechanical sensitivity norm
        $\|\nabla_{\vect{\phi}}c\|_{\mathrm{rms}}$ yields an order-of-magnitude
        initialization guide:
        \begin{equation}\label{eq:beta_scaling}
        \begin{aligned}
        \|w_h\rho_{h,0}\mat{H}^\top\mat{M}^{-1}\vect{h}\|_{\mathrm{rms}}
        &\sim \mathcal{O}(w_h\rho_{h,0}\beta^2) \\
        \Longrightarrow\quad
        \beta &\sim \mathcal{O}\!\left(\sqrt{\frac{\|\nabla_{\vect{\phi}}c\|_{\mathrm{rms}}}{w_h\rho_{h,0}}}\right).
        \end{aligned}
        \end{equation}
        Equation~\eqref{eq:beta_scaling} serves only as a rough initial scaling guide:
        in later iterations, continuation
        ramping, penalty adaptation, and clamped multipliers absorb moderate
        sensitivity differences across meshes and load cases, allowing a single
        value $\beta = 0.065$ to cover all five MBB grids and the $120\times 40$
        arch bridge. Severe initial over-scaling nonetheless locks phase
        evolution prematurely, as illustrated by the $\beta = 0.150$ case in
        Table~\ref{tab:high_res_comparison}. Conversely, when switching from
        half-MBB compliance to the unit-force inverter
        (Section~\ref{subsec:inverter}), where mechanical sensitivity drops by
        roughly three orders of magnitude, Eq.~\eqref{eq:beta_scaling} explains
        why reducing $\beta$ by $\sim\sqrt{10^3}$ (from $0.065$ to $0.002$)
        restores balanced nucleation; normalizing $\nabla_{\vect{\phi}}c$ by its
        initial RMS norm is an attractive alternative across broader objectives.
\end{itemize}
Similarly, for ultra-low volume fractions ($V_0 \le 0.15$), reducing the initial
trust radius to $\texttt{move} = 0.2V_0$--$0.3V_0$ prevents early step
oscillations from severing nascent load paths (for example, $\texttt{move} =
0.02$ lowers $240\times 160$ compliance at $V_0 = 0.10$ from $272.34$ to
$236.32$ in 80 iterations).

\begin{table}[htbp]
\centering
\footnotesize
\caption{Effect of $(w,\beta)$ on the $240\times 160$ half-MBB benchmark
($V_0 = 0.30$, $p = 3.0$).}\label{tab:high_res_comparison}
\setlength{\tabcolsep}{2.4pt}
\begin{tabular*}{\columnwidth}{@{\extracolsep{\fill}}ccrrccr@{\extracolsep{\fill}}}
\toprule
$w$ & $\beta$ & Comp.\ $c$ & $c_{0/1}$ & Iter. & $M_{\mathrm{nd}}$ (\%) & $\|\vect{h}\|_*$ \\
\midrule
$2.1$ & $0.065$ & $76.234$  & $71.902$    & $92$          & $7.40\%$  & $2.50\times 10^{-3}$ \\
$4.2$ & $0.065$ & $80.078$  & $72.251$    & $85$          & $11.73\%$ & $5.60\times 10^{-3}$ \\
$4.2$ & $0.150$ & $361.364$ & $1.33\times 10^4$ & $500^\dagger$ & $59.12\%$ & $1.72\times 10^{-1}$ \\
\bottomrule
\end{tabular*}
\vspace{0.3ex}
{\scriptsize $^\dagger$Reached the 500-iteration ceiling because barrier
over-penalization kept $\|\vect{h}\|_* > 0.28\beta$, preventing move-limit
contraction and leaving disconnected islands upon binary thresholding.}
\end{table}

\section{Extensibility Beyond Minimum Compliance}\label{sec:extensions}

To demonstrate that \texttt{topAC99} functions not merely as a single-benchmark
compliance script but as an adaptable engineering optimization kernel, we present
two distinct extensions that require zero modifications to the Kronecker
Laplacian, \texttt{conv2} mappings, AC-ALM update, or 18-line 1D-dual MMA solver.

\subsection{Compliant Mechanism Synthesis: Symmetric Force Inverter}%
\label{subsec:inverter}

Compliant mechanisms transfer motion and force through elastic member
flexure~\cite{Sigmund1997a,Bendsoe2003,ZhuBenliang2020}.
Unlike self-adjoint compliance minimization where $\partial c/\partial \bar{\phi}_e
\le 0$ everywhere, maximizing output displacement in a force inverter yields
signed mutual-strain-energy sensitivities, making it a stringent test of both the
inlined 1D-dual MMA solver and Allen--Cahn regularization.

\begin{figure*}[t]
\centering
\begin{subfigure}[b]{0.26\textwidth}
    \centering
    \resizebox{\textwidth}{!}{%
\begin{tikzpicture}[>=Stealth, every node/.style={font=\small},
    spring/.style={thick, black!80}]

    \fill[gray!6] (0, 2.6) rectangle (5.2, 4.4);
    \draw[dashed, gray!55] (0, 2.6) -- (0, 4.4) -- (5.2, 4.4) -- (5.2, 2.6);
    \node[font=\scriptsize\itshape, text=black!45] at (2.6, 3.5)
        {Mirrored upper half ($u_y$ symmetric)};

    \fill[blue!4!white] (0, 0) rectangle (5.2, 2.6);
    \draw[thick, black!85] (0, 0) rectangle (5.2, 2.6);

    \draw[dash pattern=on 6pt off 2.5pt on 1.5pt off 2.5pt, thick, blue!65!black]
        (-0.35, 2.6) -- (5.55, 2.6);
    \foreach \x in {0.8, 1.8, 2.8, 3.8, 4.4} {
        \draw[thin, blue!65!black] (\x, 2.69) circle (0.055);
    }
    \node[anchor=south, font=\scriptsize\bfseries, text=blue!65!black]
        at (2.6, 2.72) {Symmetry axis ($u_y = 0,\; \nabla\phi\cdot\vect{n} = 0$)};

    \node[font=\normalsize\bfseries, text=black!50] at (2.6, 1.30)
        {$\Omega_{\mathrm{half}}$ ($120 \times 60$)};

    \draw[ultra thick, black!90] (0, 0) -- (0, 0.35);
    \foreach \y in {0.0, 0.11, 0.22, 0.33} {
        \draw[line width=0.5pt, black!75] (0, \y) -- (-0.20, \y - 0.15);
    }
    \node[anchor=east, font=\scriptsize, text=black!80] at (-0.20, 0.16)
        {$u_x=u_y=0$};

    \draw[thick, gray!55] (0, 4.05) -- (0, 4.40);
    \foreach \y in {4.07, 4.18, 4.29, 4.40} {
        \draw[line width=0.4pt, gray!55] (0, \y) -- (-0.18, \y + 0.13);
    }

    \draw[thick, black!85] (-1.25, 2.38) -- (-1.25, 2.82);
    \foreach \y in {2.40, 2.52, 2.64, 2.76} {
        \draw[line width=0.45pt, black!70] (-1.25, \y) -- (-1.40, \y - 0.12);
    }
    \draw[spring] (-1.25, 2.6) -- (-1.05, 2.6)
        -- (-0.97, 2.73) -- (-0.83, 2.47) -- (-0.69, 2.73)
        -- (-0.55, 2.47) -- (-0.41, 2.73) -- (-0.33, 2.6) -- (0, 2.6);
    \node[anchor=south, font=\scriptsize, text=black!85] at (-0.72, 2.76)
        {$k_{\mathrm{in}}$};
    \fill[red!80!black] (0, 2.6) circle (1.8pt);
    \draw[ultra thick, -{Stealth[length=2.8mm, width=1.9mm]}, red!80!black]
        (0.04, 2.6) -- (0.85, 2.6);
    \node[anchor=north west, font=\scriptsize\bfseries, text=red!80!black]
        at (0.08, 2.55) {$f_{\mathrm{in}} = 1$};

    \fill[teal!75!black] (5.2, 2.6) circle (1.8pt);
    \draw[ultra thick, -{Stealth[length=2.8mm, width=1.9mm]}, teal!75!black]
        (5.16, 2.6) -- (4.35, 2.6);
    \node[anchor=north east, font=\scriptsize\bfseries, text=teal!75!black]
        at (5.12, 2.55) {$u_{\mathrm{out}}$};
    \draw[spring] (5.2, 2.6) -- (5.45, 2.6)
        -- (5.53, 2.73) -- (5.67, 2.47) -- (5.81, 2.73)
        -- (5.95, 2.47) -- (6.09, 2.73) -- (6.17, 2.6) -- (6.40, 2.6);
    \draw[thick, black!85] (6.40, 2.38) -- (6.40, 2.82);
    \foreach \y in {2.40, 2.52, 2.64, 2.76} {
        \draw[line width=0.45pt, black!70] (6.40, \y) -- (6.55, \y - 0.12);
    }
    \node[anchor=south, font=\scriptsize, text=black!85] at (5.82, 2.76)
        {$k_{\mathrm{out}}$};

    \draw[<->, thin, black!70] (0, -0.28) -- (5.2, -0.28)
        node[midway, fill=white, inner sep=1.5pt, font=\scriptsize] {$L = 120$};
    \draw[<->, thin, black!70] (5.55, 0) -- (5.55, 2.6)
        node[midway, fill=white, inner sep=1.5pt, font=\scriptsize, rotate=90]
        {$H = 60$};

\end{tikzpicture}}
    \caption{Half-domain schematic}
    \label{fig:inverter_setup}
\end{subfigure}
\hfill
\begin{subfigure}[b]{0.23\textwidth}
    \centering
    \includegraphics[width=\textwidth]{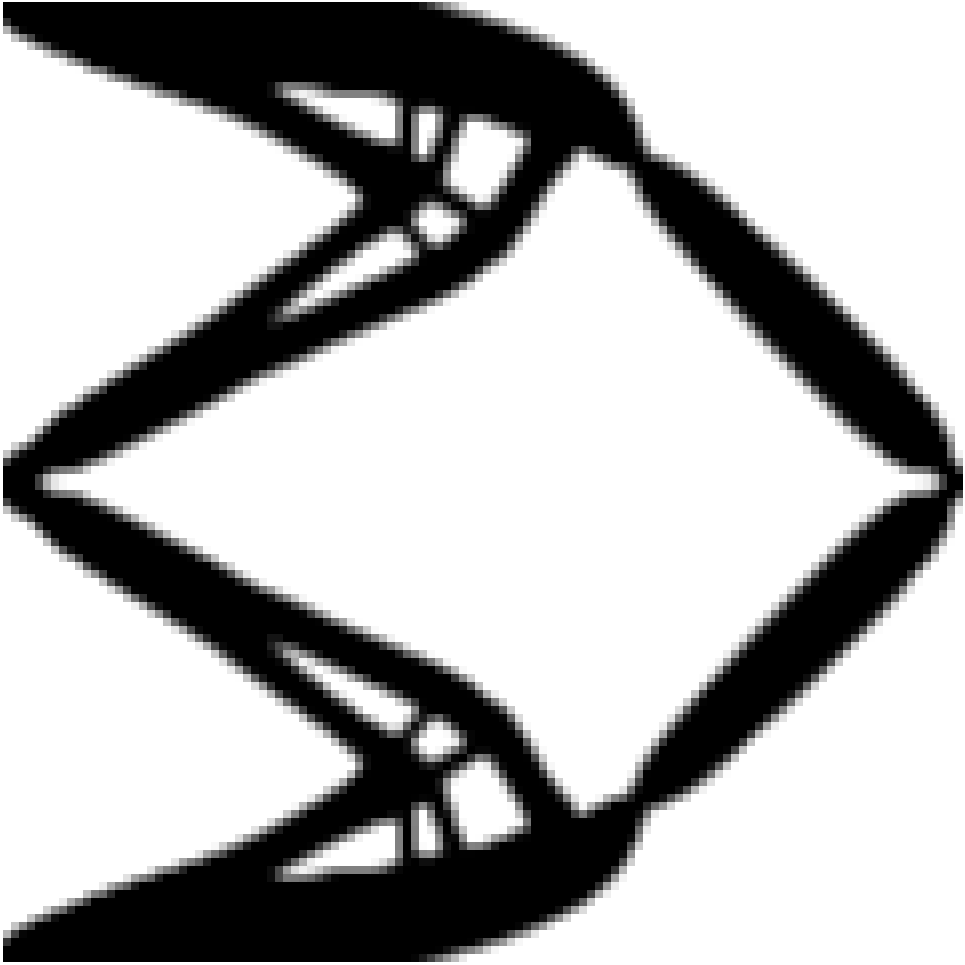}
    \caption{$120\times 60$ ($w=2.1$)}
    \label{fig:inverter_topo}
\end{subfigure}
\hfill
\begin{subfigure}[b]{0.23\textwidth}
    \centering
    \includegraphics[width=\textwidth]{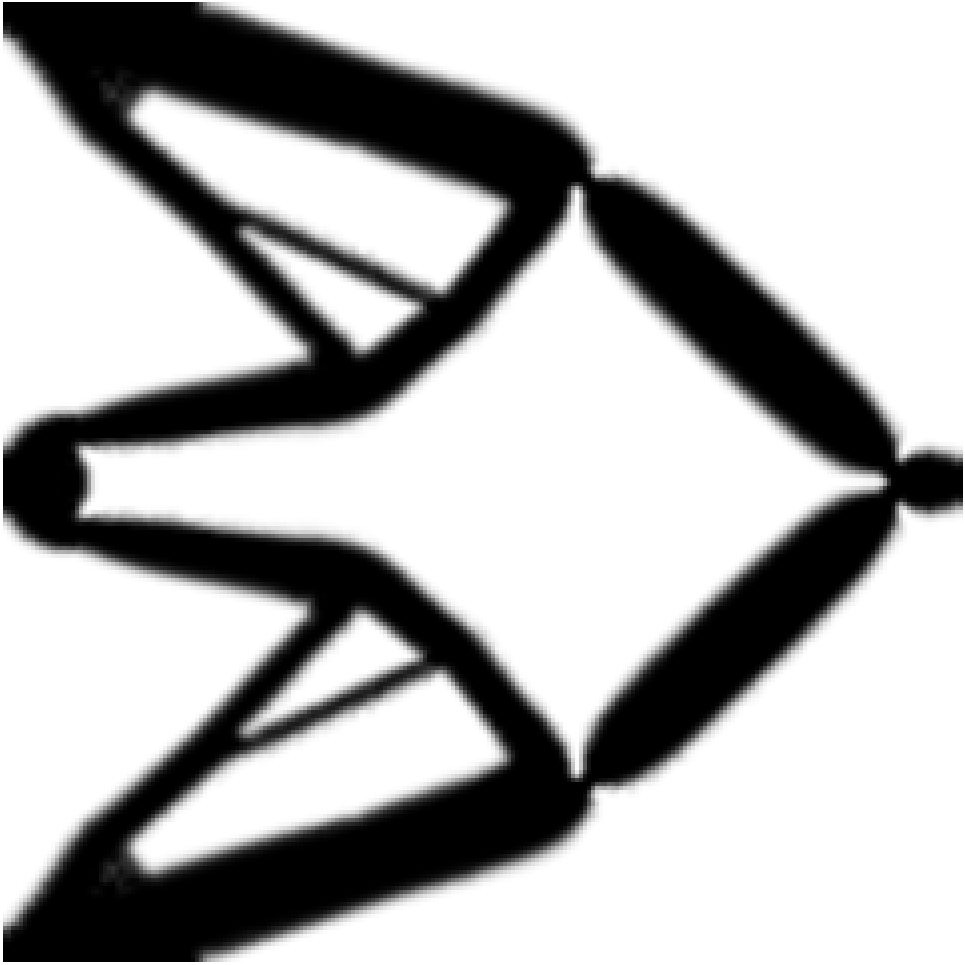}
    \caption{$240\times 120$ ($w=4.2$)}
    \label{fig:inverter_topo_w42}
\end{subfigure}
\hfill
\begin{subfigure}[b]{0.24\textwidth}
    \centering
    \includegraphics[width=\textwidth]{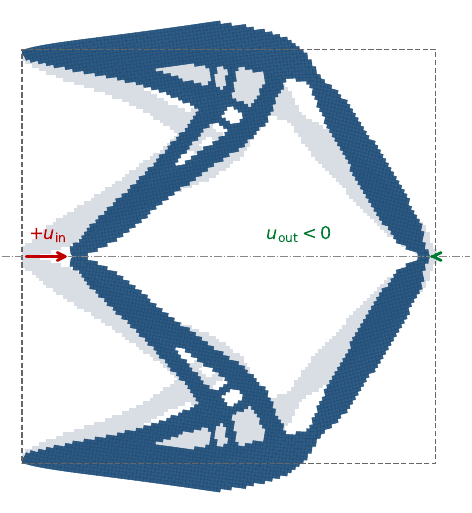}
    \caption{Deformed ($\times 1.65$)}
    \label{fig:inverter_def}
\end{subfigure}
\caption{Symmetric force inverter compliant mechanism synthesized by
\texttt{topAC99} ($V_0 = 0.30$, $\beta = 0.002$): (a)~computational half-domain
$\Omega_{\mathrm{half}}$ with port springs $k_{\mathrm{in}} = k_{\mathrm{out}} = 0.1$;
(b)~mirrored full-mechanism layout computed on the $120\times 60$ half-grid ($w = 2.1$,
$u_{\mathrm{out},0/1} = -1.0606$, 93 iterations); (c)~mesh-refined mirrored layout
computed on the $240\times 120$ half-grid at matched physical interface ratio
$w/L_x = 0.0175$ ($w = 4.2$, $u_{\mathrm{out},0/1} = -0.9999$);
(d)~volume-preserving $0/1$ undeformed vs.\ amplified deformed silhouette ($\times 1.65$)
on the $120\times 60$ half-grid.}\label{fig:inverter}
\end{figure*}
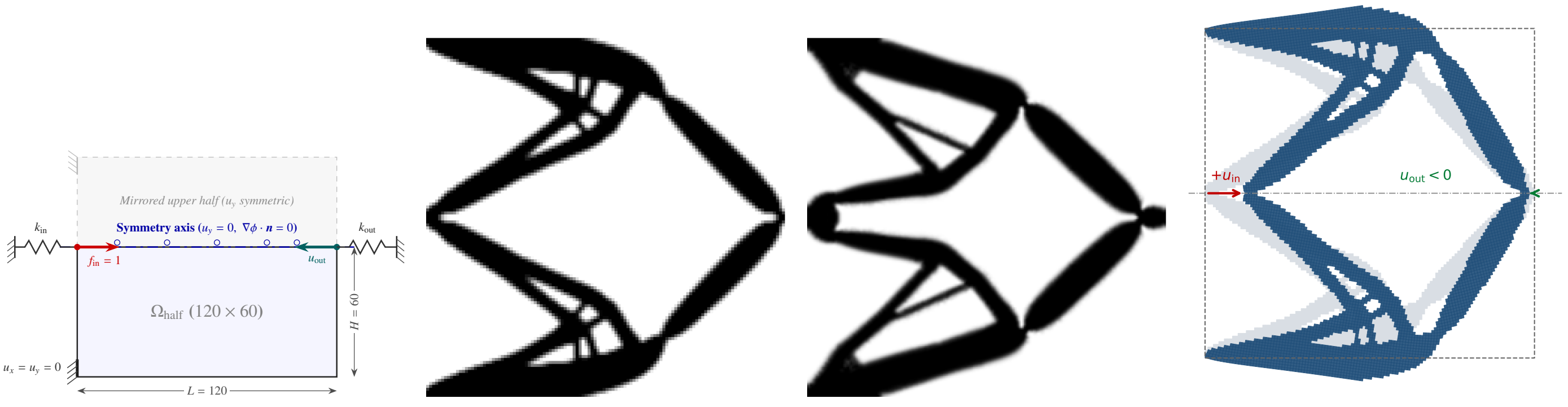

We consider the canonical force inverter benchmark~\cite{Bendsoe2003,WangFengwen2011a}
modeled on its lower symmetric half-domain $\Omega_{\mathrm{half}}$ of size
$n_x \times n_y$ with aspect ratio $2:1$ ($V_0 = 0.30$, $p = 3.0$), as depicted
in Fig.~\ref{fig:inverter}a. Along the top horizontal symmetry axis ($y = H$,
grid row $1$), vertical displacements vanish ($u_y = 0$). Image-filtering codes such as \texttt{top99neo}~\cite{Ferrari2020b} require
explicit boundary-padding switches (\texttt{bcF = 'symmetric'}) to avoid
density-filter distortion across symmetry planes. By contrast,
\texttt{topAC99}'s Kronecker Laplacian \texttt{Klap} enforces homogeneous
Neumann zero-flux conditions $\nabla\phi\cdot\vect{n} = 0$ on
$\partial\Omega_{\mathrm{half}}$ automatically, guaranteeing $C^1$-symmetric
phase profiles across the symmetry axis with zero code changes.

The bottom element on the left edge is clamped ($u_x = u_y = 0$). On
$\Omega_{\mathrm{half}}$, a rightward actuation force $f_{\mathrm{in}} = 1$
with actuator spring stiffness $k_{\mathrm{in}} = 0.1$ acts at the top-left
input port $d_{\mathrm{in}} = 1$. At the top-right output port $d_{\mathrm{out}}
= 2n_x(n_y+1) + 1$, a grounded workpiece spring $k_{\mathrm{out}} = 0.1$
resists motion (corresponding by symmetry to $2f_{\mathrm{in}}$ and
$2k_{\mathrm{in}}, 2k_{\mathrm{out}}$ on the full
mechanism~\cite{Bendsoe2003,WangFengwen2011a}). Let
$\mat{K}_{\mathrm{tot}}(\bar{\vect{\phi}}) = \mat{K}(\bar{\vect{\phi}}) +
k_{\mathrm{in}}\vect{e}_{d_{\mathrm{in}}}\vect{e}_{d_{\mathrm{in}}}^\top +
k_{\mathrm{out}}\vect{e}_{d_{\mathrm{out}}}\vect{e}_{d_{\mathrm{out}}}^\top$.
Solving the two-column linear system $\mat{K}_{\mathrm{tot}}[\vect{u}_1,\vect{u}_2]
= [\vect{e}_{d_{\mathrm{in}}}, \vect{e}_{d_{\mathrm{out}}}]$ simultaneously for
the physical actuation load $\vect{f}_1 = \vect{e}_{d_{\mathrm{in}}}$ and unit
virtual output load $\vect{f}_2 = \vect{e}_{d_{\mathrm{out}}}$, the objective to
minimize is the output displacement $c(\bar{\vect{\phi}}) = u_{\mathrm{out}} =
\vect{e}_{d_{\mathrm{out}}}^\top\vect{u}_1$ (driving $u_{\mathrm{out}} < 0$
leftward). Differentiating $\vect{f}_2^\top\vect{u}_1$ via the adjoint relation
$\mat{K}_{\mathrm{tot}}\vect{\psi} = -\vect{f}_2 \implies \vect{\psi} = -\vect{u}_2$
yields the exact elemental sensitivity:
\begin{equation}\label{eq:inverter_sens}
\frac{\partial c}{\partial \bar{\phi}_e}
= -\vect{u}_{e,2}^\top \frac{\partial \mat{k}_e}{\partial \bar{\phi}_e}\vect{u}_{e,1}
= -p(E_0 - E_{\min})\bar{\phi}_e^{p-1}\bigl(\vect{u}_{e,1}^\top\mat{k}_0\vect{u}_{e,2}\bigr).
\end{equation}
Because the mutual energy density $\texttt{ce}_e = \vect{u}_{e,1}^\top\mat{k}_0\vect{u}_{e,2}$
takes both positive and negative values across elements, Line~42 of
\texttt{topAC99.m} (\texttt{dc = -penal*(E0-Emin)*x.\^{}(penal-1).*ce}) remains
completely unchanged. Adapting \texttt{topAC99.m} to the force inverter requires
replacing only Lines~17--19 and Lines~38--41 (7 kernel lines replaced by 9 lines):
\begin{lstlisting}[numbers=none]
% Force Inverter Ports, Loads & Symmetry BCs (Replaces Lines 17--19)
din = 1; dout = 2*nelx*(nely+1) + 1;
U = zeros(2*(nely+1)*(nelx+1), 2);
F = sparse([din, dout], [1, 2], [1, 1], 2*(nely+1)*(nelx+1), 2);
fixeddofs = [2*(nely+1)+(-3:0), 2:2*(nely+1):2*(nelx+1)*(nely+1)];
\end{lstlisting}
\begin{lstlisting}[numbers=none]
% Springs, Two-Column Solve & Mutual Energy (Replaces Lines 38--41)
K = sparse([iK; din; dout], [jK; din; dout], [sK; 0.1; 0.1]); K = (K + K')/2;
U(freedofs,:) = K(freedofs, freedofs) \ F(freedofs,:);
U1 = U(:,1); U2 = U(:,2);
ce = reshape(sum((U1(edofMat)*KE).*U2(edofMat), 2), nely, nelx);
c = U(dout, 1);
\end{lstlisting}

Guided by Eq.~\eqref{eq:beta_scaling}, setting $(w, \beta) = (2.1, 0.002)$ on
a $120\times 60$ half-grid converges smoothly in 93 iterations to an inverted
output displacement of $u_{\mathrm{out}} = -1.0180$ ($u_{\mathrm{in}} =
8.8772$, geometric advantage $\mathrm{GA} = |u_{\mathrm{out}}|/u_{\mathrm{in}}
= 0.115$). Strict volume-preserving $0/1$ thresholding yields an even stronger
stroke $u_{\mathrm{out},0/1} = -1.0606$ ($\mathrm{GA}_{0/1} = 0.121$),
confirming that motion inversion is carried by solid elastic flexure without
relying on intermediate-density hinges. At convergence, $\|\vect{h}\|_* =
3.69\times 10^{-4} \le 0.28\beta$ and $|V - V_0| = 1.17\times 10^{-4}$.

Table~\ref{tab:inverter_scaling} evaluates the force inverter across four grid
resolutions ($80\times 40$ to $240\times 120$) at fixed element width $w = 2.1$.
It also includes a fixed-\emph{physical}-width refinement run ($240\times 120,
w = 4.2$, matching $w/L_x = 0.0175$ of $120\times 60, w = 2.1$), a
variance-matched density-filtered SIMP+MMA baseline ($r_{\min} =
2\sqrt{3}\,\ell_{\mathrm{AC}} = 2.34$), and an unregularized nodal baseline
($w = 0, \beta = 0$). Three numerical findings stand out:
\begin{enumerate}
    \item \textit{Escape from the sign-reversal saddle point}: From the uniform
        initial state ($\bar{\phi}_e = 0.30$), rightward actuation stretches the
        continuum so $u_{\mathrm{out}} = +0.1014 > 0$ initially, crossing zero
        near iterations 6--30. Thanks to the zero-crossing denominator guard on
        Line~57 and the $\|\vect{h}\|_* \le 0.28\beta$ stationarity gate,
        \texttt{topAC99} traverses $c \approx 0$ without premature move-limit
        contraction, whereas the unregularized baseline ($w = 0, \beta = 0$)
        stalls at $u_{\mathrm{out}} = 0.0000$ ($M_{\mathrm{nd}} = 65.40\%$).
    \item \textit{Sharper phase contrast than variance-matched density filtering}:
        On $120\times 60$ at matched second-moment length scale ($w = 2.1
        \leftrightarrow r_{\min} = 2.34$) and MMA move-limit annealing, both
        formulations require nearly identical runtime and iteration counts.
        However, \texttt{topAC99} achieves stronger raw and thresholded stroke
        inversion while reducing non-discreteness $M_{\mathrm{nd}}$ by $30\%$
        ($8.19\%$ vs.\ $11.61\%$).
    \item \textit{Fixed-element vs.\ fixed-physical width and hinge resolution}:
        At fixed $w = 2.1$, iteration counts remain tightly bounded ($83$--$95$)
        and $M_{\mathrm{nd}}$ drops to $4.44\%$ on $240\times 120$. Scaling $w = 4.2$
        proportionally on $240\times 120$ (Fig.~\ref{fig:inverter}c) suppresses
        secondary web sub-splitting while preserving thresholded stroke within
        $0.15\%$ ($u_{\mathrm{out},0/1} = -0.9999$ vs.\ $-1.0014$), albeit requiring
        316 iterations as diffusive coarsening across a $4.2$-element band slows
        under contracted move limits. As expected for any single-field formulation
        without multi-field dilated/eroded projections~\cite{WangFengwen2011a,Lazarov2016a},
        maximizing stroke under linear kinematics still favors localized flexural
        pivots ($1$--$2$ elements wide); coupling \texttt{topAC99} with stress
        or robust erosion--dilation constraints is a natural avenue for future work.
\end{enumerate}

\begin{table*}[t]
\centering
\footnotesize
\caption{Multi-mesh scaling, physical-width consistency, and baseline comparison
on the symmetric force inverter ($V_0 = 0.30$, $p = 3.0$, $k_{\mathrm{in}} = k_{\mathrm{out}} = 0.1$).}\label{tab:inverter_scaling}
\setlength{\tabcolsep}{4.2pt}
\begin{tabular*}{\textwidth}{@{\extracolsep{\fill}}llccccrrrr@{\extracolsep{\fill}}}
\toprule
Half-Grid & Formulation & $u_{\mathrm{out}}$ & $u_{\mathrm{out},0/1}$ & $\mathrm{GA}$ ($\mathrm{GA}_{0/1}$) & Iter. & $M_{\mathrm{nd}}$ (\%) & $\|\vect{h}\|_*$ & $|V - V_0|$ & Time (s) \\
\midrule
$80 \times 40$   & \texttt{topAC99} ($w=2.1,\,\beta=0.002$) & $-1.0462$ & $-1.0979$ & $0.118$ ($0.125$) & $83$  & $8.30\%$  & $4.73\times 10^{-4}$ & $9.85\times 10^{-5}$ & $1.35$ \\
$120 \times 60$  & \texttt{topAC99} ($w=2.1,\,\beta=0.002$) & $-1.0180$ & $-1.0606$ & $0.115$ ($0.121$) & $93$  & $8.19\%$  & $3.69\times 10^{-4}$ & $1.17\times 10^{-4}$ & $3.58$ \\
$180 \times 90$  & \texttt{topAC99} ($w=2.1,\,\beta=0.002$) & $-0.9788$ & $-1.0091$ & $0.110$ ($0.115$) & $95$  & $6.13\%$  & $2.64\times 10^{-4}$ & $1.55\times 10^{-5}$ & $9.37$ \\
$240 \times 120$ & \texttt{topAC99} ($w=2.1,\,\beta=0.002$) & $-0.9781$ & $-1.0014$ & $0.110$ ($0.113$) & $94$  & $4.44\%$  & $2.00\times 10^{-4}$ & $2.77\times 10^{-6}$ & $18.46$ \\
$240 \times 120$ & \texttt{topAC99} ($w=4.2,\,\beta=0.002$) & $-0.9414$ & $-0.9999$ & $0.103$ ($0.111$) & $316$ & $11.04\%$ & $1.79\times 10^{-4}$ & $1.29\times 10^{-4}$ & $63.67$ \\
\midrule
$120 \times 60$  & Filtered SIMP + MMA ($r_{\min}=2.34$)    & $-0.9826$ & $-1.0522$ & $0.110$ ($0.120$) & $88$  & $11.61\%$ & ---                  & $4.43\times 10^{-5}$ & $3.46$ \\
$240 \times 120$ & Unregularized nodal ($w=0,\,\beta=0$)    & $+0.0000^\ddagger$ & $+0.0000$ & $0.000$ ($0.000$) & $30$ & $65.40\%$ & ---       & $5.82\times 10^{-2}$ & $5.20$ \\
\bottomrule
\end{tabular*}
\vspace{0.3ex}
{\scriptsize $^\ddagger$Prematurely satisfied $\text{ch} < 0.01$ at the
zero-displacement saddle point without double-well phase segregation.}
\end{table*}

\subsection{Civil Structural Design: Highway Deck Arch Bridge with Passive Solid Deck}%
\label{subsec:bridge}

\begin{figure}[htbp]
\centering
\begin{subfigure}[b]{\columnwidth}
    \centering
    \resizebox{0.94\columnwidth}{!}{%
\begin{tikzpicture}[>=Stealth, every node/.style={font=\small}]

    \fill[blue!4!white] (0, 0) rectangle (6.6, 2.2);
    \draw[thick, black!85] (0, 0) rectangle (6.6, 2.2);

    \fill[black!30] (0, 1.98) rectangle (6.6, 2.2);
    \draw[thick, black!85] (0, 1.98) -- (6.6, 1.98);
    \node[font=\scriptsize\bfseries, text=black!95] at (3.3, 2.09)
        {Passive solid roadway deck ($\phi = 1$)};

    \node[font=\large\bfseries, text=black!45] at (3.3, 1.10)
        {$\Omega$ (Designable domain)};

    \draw[<->, thin, black!70] (0, 0.35) -- (6.6, 0.35)
        node[midway, fill=blue!4!white, inner sep=1.5pt, font=\scriptsize] {$L = 120$};

    \draw[thick, red!80!black] (0, 2.7) -- (6.6, 2.7);
    \foreach \i in {0, 1, ..., 15} {
        \draw[-{Stealth[length=2.2mm, width=1.4mm]}, thick, red!80!black]
            ({\i*6.6/15}, 2.7) -- ({\i*6.6/15}, 2.2);
    }
    \node[anchor=south, font=\scriptsize\bfseries, text=red!80!black]
        at (3.3, 2.72) {Distributed traffic load $q$};

    \fill[black!70] (0, 0) -- (-0.20, -0.30) -- (0.20, -0.30) -- cycle;
    \draw[thick, black!85] (-0.32, -0.30) -- (0.32, -0.30);
    \fill[pattern=north east lines, pattern color=black!50]
        (-0.32, -0.45) rectangle (0.32, -0.30);

    \fill[black!70] (6.6, 0) -- (6.40, -0.30) -- (6.80, -0.30) -- cycle;
    \draw[thick, black!85] (6.28, -0.30) -- (6.92, -0.30);
    \fill[pattern=north east lines, pattern color=black!50]
        (6.28, -0.45) rectangle (6.92, -0.30);

    \draw[<->, thin, black!70] (7.10, 0) -- (7.10, 2.2)
        node[midway, fill=white, inner sep=1.5pt, font=\scriptsize, rotate=90] {$H = 40$};

\end{tikzpicture}}
    \caption{Problem schematic and passive solid roadway deck}
    \label{fig:bridge_setup}
\end{subfigure}
\vskip 0.4em
\begin{subfigure}[b]{\columnwidth}
    \centering
    \includegraphics[width=0.94\columnwidth]{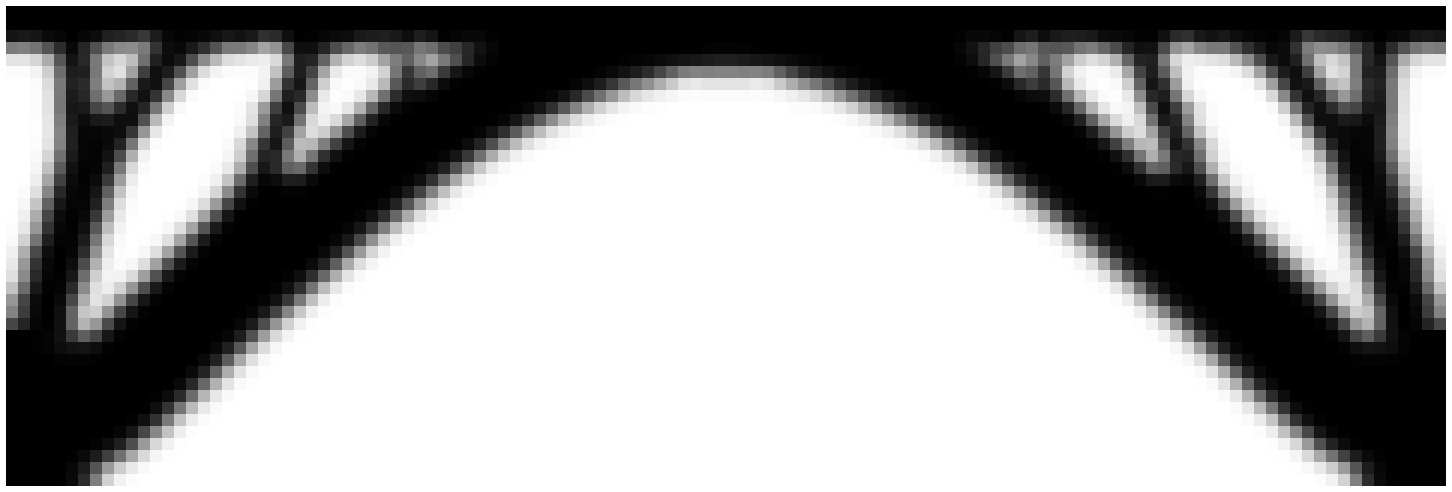}
    \caption{Optimized arch bridge topology ($c = 6.501$, $c_{0/1} = 5.938$, 87 iters)}
    \label{fig:bridge_topo}
\end{subfigure}
\caption{Highway deck arch bridge with passive solid deck on a $120\times 40$
domain ($V_0 = 0.40$, $w = 2.1$, $\beta = 0.065$).}\label{fig:bridge}
\end{figure}
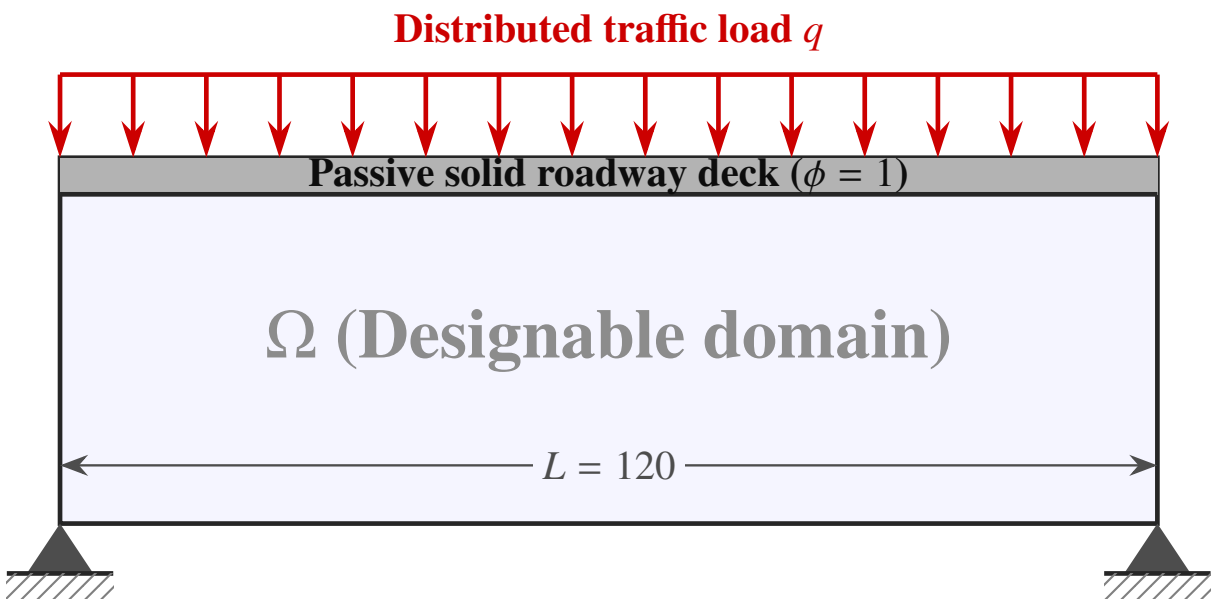
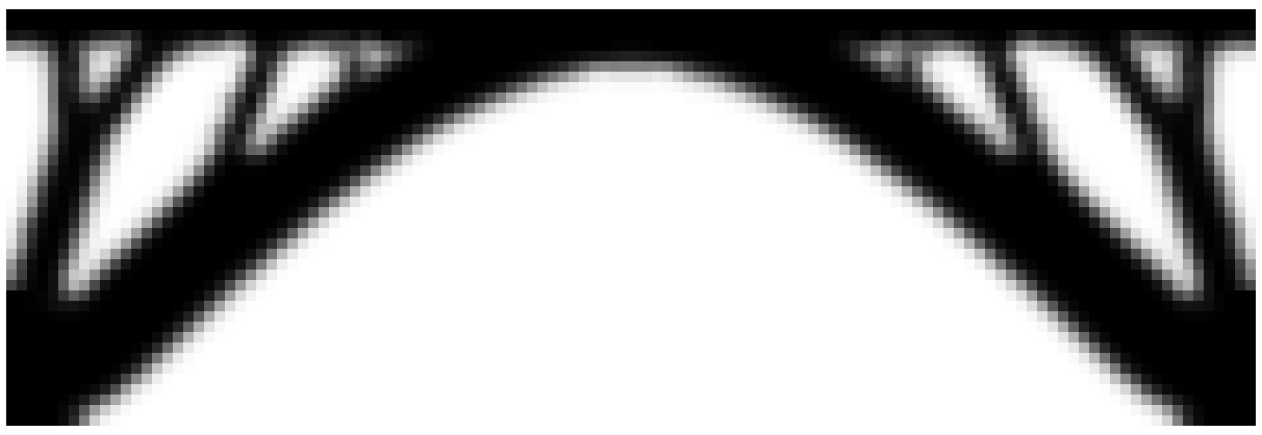

Practical structural engineering applications frequently combine distributed
service loads with non-designable passive subdomains. Figure~\ref{fig:bridge}a
models a $120\times 40$ highway deck arch bridge ($V_0 = 0.40$, aspect ratio
$3:1$) carrying a uniform downward traffic load $q$ across a continuous
two-element passive solid roadway deck (nodal rows $1:3$), supported by pinned
bedrock abutments ($u_x = u_y = 0$) at both bottom corners $(0,0)$ and $(L,0)$.

In \texttt{topAC99.m}, passive \emph{solid} subdomains are enforced at the
nodal level by initializing the passive deck nodes to $\phi_i = 1$ after
Line~30 and clamping \texttt{a\_lim} and \texttt{b\_lim} after Line~72 without
per-iteration \texttt{meshgrid} overhead. Initializing $\phi_i = 1$ prior to
\texttt{upp = phi + 0.5} guarantees $\phi_i^{(1)} = 1 < U_i^{(1)} = 1.5$,
preventing upper MMA asymptote crossings for any $V_0 \in (0,1)$, while
distributed deck loads and pinned abutments replace Lines~18--19:
\begin{lstlisting}[numbers=none]
% Distributed Deck Load & Bedrock Pins (Replaces Lines 18--19)
deck_nodes = 1:(nely+1):(nelx+1)*(nely+1);
f_deck = [0.5; ones(nelx-1,1); 0.5] / nelx;
F = sparse(2*deck_nodes, 1, -f_deck, 2*(nely+1)*(nelx+1), 1);
fixeddofs = [2*(nely+1)+(-1:0), 2*(nelx+1)*(nely+1)+(-1:0)];
\end{lstlisting}
\begin{lstlisting}[numbers=none]
% Passive Solid Deck Initialization & Clamping
phi(1:3, :) = 1.0; p_old1 = phi; p_old2 = phi; % after Line 30
a_lim(1:3, :) = 1.0; b_lim(1:3, :) = 1.0;      % after Line 72
\end{lstlisting}
Here \texttt{f\_deck} assigns work-equivalent trapezoidal nodal weights summing
to unit downward load. (For general passive \emph{void} elements bordering active
solid nodes, one additionally sets \texttt{x(pasV) = 0; dc(pasV) = 0} after
Lines~35 and~42 so shared boundary nodes do not leak $O((0.25)^p)$ stiffness into
void cells.) Because the 1D-dual MMA surrogate retains clamped deck nodes inside
\texttt{r1} and \texttt{p1}, total material volume (passive deck plus substructure)
satisfies $V \le V_0$ automatically without active-set re-indexing.

Executing this modified script with $(w,\beta) = (2.1, 0.065)$ converges in 87
iterations ($2.15$\,s) to $c = 6.501$ and $c_{0/1} = 5.938$ ($|V - V_0| =
6.36\times 10^{-5}$, Fig.~\ref{fig:bridge}b), generating a parabolic compression
arch anchored into the bedrock abutments with evenly spaced spandrel columns. For
reference, under the same $\text{ch} < 0.01$ stopping threshold, a
variance-matched density-filtered SIMP+OC baseline ($r_{\min} = 2.34$) exhibits
comparable FEA-governed per-step cost ($28.1$\,ms vs.\ $24.7$\,ms/iter) but
takes 191 fixed-move iterations ($5.37$\,s) to reach an equivalent thresholded
compliance $c_{0/1} = 5.902$ ($c = 6.383$, $M_{\mathrm{nd}} = 15.29\%$). Because this minimal-line implementation clamps rows \texttt{1:3}
(\texttt{a\_lim = b\_lim = 1}) while evaluating \texttt{Klap*phi(:)} and
$\vect{h}(\vect{\phi})$ across all $\Nn$ nodes, two implementation-specific
boundary effects arise. First, Laplacian coupling across clamped row~3
produces a diffuse transition halo of thickness $\sim w$ beneath unloaded deck
segments ($M_{\mathrm{nd}} = 17.43\%$). Second, the full-domain residual norm
$\|\vect{h}\|_* = 3.49\times 10^{-3} \le 0.28\beta$ includes nonzero
$-\epsilon^2\Delta\phi$ reactions localized on the clamped passive boundary
nodes; restricting $\vect{h}$ and $\vect{\lambda}_h$ to free nodes or
decoupling \texttt{Klap} across passive interfaces removes this boundary
contribution.

\section{Conclusions}\label{sec:concl}

This paper presented \texttt{topAC99}, a self-contained 99-line MATLAB software
kernel for filter-free Allen--Cahn topology optimization via the Augmented
Lagrangian Method. By reformulating the continuum steady-state Allen--Cahn residual-constrained
framework~\cite{DuRun2026topac} specifically for structured grid
efficiency, \texttt{topAC99} eliminates the multi-point Gauss quadrature loops,
explicit sparse filter/transfer matrices, and external optimization libraries
required by full finite element research codes.

The software architecture combines three discrete building blocks: (i)~a
lumped tensor-product discretization with a 7-line Kronecker-sum 5-point
Laplacian that enforces zero-flux Neumann conditions at width-invariant
$\mathcal{O}(\Nn)$ memory cost; (ii)~matrix-free primal density averaging and
adjoint sensitivity backpropagation via 2D convolutions (\texttt{conv2}); and
(iii)~an inlined 18-line 1D-dual bisection MMA solver for signed sensitivities
with conservative volume feasibility. Benchmarks up to $121{,}002$ elastic DOFs
confirm that Allen--Cahn regularization adds $<1.2\%$ per-step overhead onto
FEA-dominated iterations and converges within $78$--$95$ iterations on fine
meshes. An order-of-magnitude initialization guide for $\beta$ further
streamlines adaptation across different objective scales. Minimal-line
extensions to a symmetric force inverter and a passive-deck arch bridge
demonstrate that \texttt{topAC99} provides a compact, extensible, and
dependency-free foundation for Allen--Cahn topology optimization.

\section*{CRediT authorship contribution statement}
\textbf{Run Du:} Conceptualization, Methodology, Software, Validation, Formal
analysis, Investigation, Visualization, Writing -- original draft, Writing --
review \& editing, Funding acquisition.

\section*{Declaration of competing interest}
The author declares that there are no known competing financial interests or
personal relationships that could have appeared to influence the work reported in
this paper.

\section*{Data and code availability}
The complete 99-line MATLAB implementation, \texttt{topAC99.m}, is provided in
\ref{sec:appendix}. The code is self-contained and requires no external
toolboxes or optimization packages.

Implementation details for the compliant force-inverter and arch-bridge
extensions are presented in Section~\ref{sec:extensions}. The corresponding
standalone scripts, benchmark drivers, and post-processing routines are
available from the corresponding author upon reasonable request, including for
editorial and peer-review verification.

\section*{Acknowledgments}
This work was supported by the National Natural Science Foundation of China
(Grant No.\ 51405397), the Sichuan Science and Technology Program (Grant No.\
2023YFG0182), and the Southwest Jiaotong University Multidisciplinary Research
Fund (Grant No.\ 2682025ZD005).

\onecolumn
\appendix
\section{Core MATLAB Kernel (\texorpdfstring{\texttt{topAC99.m}}{topAC99.m})}\label{sec:appendix}

\begin{lstlisting}[language=Matlab,
    firstline=1,
    lastline=99,
    basicstyle=\ttfamily\fontsize{6pt}{6.35pt}\selectfont,
    caption={\texttt{topAC99.m}: Self-contained 99-line MATLAB kernel for
    filter-free Allen--Cahn topology optimization.}]
function [c, x, loop, h_norm] = topAC99(nelx, nely, volfrac, penal, w, beta)
%% 0. PHYSICAL INTERFACE PARAMETERS
ell = w / (sqrt(2)*log(9));
ep = ell * sqrt(beta);
%% 1. MATERIAL & FINITE ELEMENT SETUP
E0 = 1.0; Emin = 1e-9; nu = 0.3;
A11 = [12  3 -6 -3;  3 12  3  0; -6  3 12 -3; -3  0 -3 12];
A12 = [-6 -3  0  3; -3 -6 -3 -6;  0 -3 -6  3;  3 -6  3 -6];
B11 = [-4  3 -2  9;  3 -4 -9  4; -2 -9 -4 -3;  9  4 -3 -4];
B12 = [ 2 -3  4 -9; -3  2  9 -2;  4  9  2  3; -9 -2  3  2];
KE = 1/(1-nu^2)/24*([A11 A12; A12' A11] + nu*[B11 B12; B12' B11]);
nodenrs = reshape(1:(nelx+1)*(nely+1), nely+1, nelx+1);
edofVec = reshape(2*nodenrs(1:end-1,1:end-1)+1, nelx*nely, 1);
edofMat = repmat(edofVec,1,8) + repmat([0 1 2*nely+[2 3 0 1] -2 -1], nelx*nely, 1);
iK = reshape(kron(edofMat, ones(8,1))', 64*nelx*nely, 1);
jK = reshape(kron(edofMat, ones(1,8))', 64*nelx*nely, 1);
U = zeros(2*(nely+1)*(nelx+1), 1);
F = sparse(2, 1, -1, 2*(nely+1)*(nelx+1), 1);
fixeddofs = [1:2:2*(nely+1), 2*(nelx+1)*(nely+1)];
freedofs = setdiff(1:2*(nely+1)*(nelx+1), fixeddofs);
%% 2. CONFORMING LAPLACIAN & MASS OPERATORS
Dx = spdiags([-ones(nelx+1,1) 2*ones(nelx+1,1) -ones(nelx+1,1)], -1:1, nelx+1, nelx+1);
Dx(1,1) = 1; Dx(end,end) = 1;
Dy = spdiags([-ones(nely+1,1) 2*ones(nely+1,1) -ones(nely+1,1)], -1:1, nely+1, nely+1);
Dy(1,1) = 1; Dy(end,end) = 1;
mx = [0.5; ones(nelx-1,1); 0.5]; my = [0.5; ones(nely-1,1); 0.5];
M = my * mx'; Minv = 1 ./ M; Omega = nelx * nely;
Klap = kron(Dx, spdiags(my, 0, nely+1, nely+1)) + kron(spdiags(mx, 0, nelx+1, nelx+1), Dy);
%% 3. INITIALIZATION
phi = repmat(volfrac, nely+1, nelx+1); p_old1 = phi; p_old2 = phi;
low = phi - 0.5; upp = phi + 0.5;
lambda_h = zeros(nely+1, nelx+1); rho_h = 0.1; c_old = 1e6; move = 0.10;
%% 4. AUGMENTED LAGRANGIAN MMA OPTIMIZATION LOOP
for loop = 1:500
    x = conv2(phi, [1 1; 1 1]/4, 'valid');
    % FE-Analysis & Sensitivity
    sK = reshape(KE(:)*(Emin + (E0 - Emin)*x(:)'.^penal), 64*nelx*nely, 1);
    K = sparse(iK, jK, sK); K = (K + K')/2;
    U(freedofs) = K(freedofs, freedofs) \ F(freedofs);
    ce = reshape(sum((U(edofMat)*KE).*U(edofMat), 2), nely, nelx);
    c = sum(sum((Emin + (E0 - Emin)*x.^penal).*ce));
    dc = -penal * (E0 - Emin) * x.^(penal-1) .* ce;
    grad_c = conv2(dc, [1 1; 1 1]/4, 'full');
    % Conserved Allen-Cahn Residual & Zero-Adjoint Gradient
    dW = beta * (2*phi - 6*phi.^2 + 4*phi.^3);
    d2W = beta * (2 - 12*phi + 12*phi.^2);
    mu = sum(M(:).*dW(:)) / Omega;
    h = ep^2 * reshape(Klap*phi(:), nely+1, nelx+1) + M.*(dW - mu);
    h_norm = sqrt(sum(h(:).^2 ./ M(:)) / Omega);
    z = lambda_h + rho_h * (h .* Minv);
    z_bar = sum(M(:).*z(:)) / Omega;
    grad_h = ep^2 * reshape(Klap*z(:), nely+1, nelx+1) + M.*d2W.*(z - z_bar);
    % Total Sensitivity & Moving Asymptotes Update
    w_h = min(1, 0.05 + 0.06 * loop);
    df0 = grad_c + w_h * grad_h;
    df1 = M / Omega;
    if loop > 30 && abs(c - c_old)/max(abs(c),1e-6) < 1e-3 && h_norm <= 0.28*beta, move = max(0.005, move*0.95); end
    if loop > 2
        gam = ones(size(phi));
        gam((phi - p_old1).*(p_old1 - p_old2) > 0) = 1.2;
        gam((phi - p_old1).*(p_old1 - p_old2) < 0) = 0.7;
        low = max(phi - 10, min(phi - 0.01, phi - gam.*(p_old1 - low)));
        upp = min(phi + 10, max(phi + 0.01, phi + gam.*(upp - p_old1)));
    end
    p_old2 = p_old1; p_old1 = phi;
    % Closed-Form 1D-Dual MMA Subproblem Solver
    p0 = (max(0, df0) + 0.001*abs(df0) + 1e-6).*(upp - phi).^2;
    q0 = (max(0, -df0) + 0.001*abs(df0) + 1e-6).*(phi - low).^2;
    p1 = df1.*(upp - phi).^2;
    r1 = sum(df1(:).*phi(:)) - volfrac - sum(p1(:)./(upp(:) - phi(:)));
    a_lim = max(0, max(phi - move, low + 0.1*(phi - low)));
    b_lim = min(1, min(phi + move, upp - 0.1*(upp - phi)));
    phi_trial = max(a_lim, min(b_lim, (sqrt(p0).*low + sqrt(q0).*upp)./(sqrt(p0) + sqrt(q0))));
    if r1 + sum(p1(:)./(upp(:) - phi_trial(:))) > 0
        l1 = 0; l2 = 1e6; phi_feas = a_lim;
        for it = 1:25
            lmid = 0.5 * (l1 + l2);
            pj = p0 + lmid * p1;
            phi_trial = max(a_lim, min(b_lim, (sqrt(pj).*low + sqrt(q0).*upp)./(sqrt(pj) + sqrt(q0))));
            if r1 + sum(p1(:)./(upp(:) - phi_trial(:))) > 0
                l1 = lmid; else, l2 = lmid; phi_feas = phi_trial;
            end
        end; phi_trial = phi_feas;
    end
    ch = max(abs(phi_trial(:) - phi(:)));
    % Display Iteration Status & Topology
    fprintf('It.:%3d  Comp.:%7.3f  ||h||_*:%7.2e  ch.:%6.4f\n', loop, c, h_norm, ch);
    colormap(gray); imagesc(1 - x); caxis([0 1]);
    axis equal; axis tight; axis off; drawnow;
    if loop >= 30 && ch < 0.01 && h_norm <= 0.28*beta, break; end
    phi = phi_trial;
    % ALM Multiplier & Penalty Schedule
    lambda_h = lambda_h + rho_h * (h .* Minv);
    lambda_h = max(-10, min(10, lambda_h));
    lambda_h = lambda_h - sum(M(:).*lambda_h(:)) / Omega;
    if mod(loop, 10) == 0 && h_norm > 0.015*beta, rho_h = min(10.0, rho_h * 1.1); end
    c_old = c;
end
end
\end{lstlisting}

\end{document}